\documentclass[twocolumn, tighten]{aastex62}
\usepackage{amsmath}
\usepackage{multirow}
\usepackage{gensymb}
\usepackage{hyperref}
\usepackage{pgfplotstable}
\usepackage{array}
\hypersetup{colorlinks, linkcolor={blue}, citecolor={blue}, urlcolor={red}}  

\shorttitle{Primordial Disk Frequencies in NGC 1333, IC 348, and Orion A}
\shortauthors{Y.-H. Yao et al.}
\submitjournal{ApJ}

\begin{document}

\title{IN-SYNC. VIII. Primordial Disk Frequencies\\
	in NGC 1333, IC 348, and the Orion A Molecular Cloud}

\author{Yuhan Yao}
\affil{Cahill Center for Astrophysics, California Institute of Technology, Pasadena, CA 91125, USA}
\affil{Department of Astronomy, Peking University, Yi He Yuan Lu 5, Hai Dian District, Beijing 100871, China}

\author{Michael R. Meyer}
\affil{Department of Astronomy, University of Michigan, Ann Arbor, MI 48109, USA}

\author{Kevin R. Covey}
\affil{Department of Physics \& Astronomy, Western Washington University, Bellingham, WA 
98225, USA}

\author{Jonathan C. Tan}
\affil{Department of Space, Earth and Environment, Chalmers University of Technology, Gothenburg, Sweden}
\affil{Department of Astronomy, University of Virginia, Charlottesville, Virginia 22904, USA}

\author{Nicola Da Rio}
\affil{Department of Astronomy, University of Virginia, Charlottesville, Virginia 22904, USA}

\begin{abstract}
In this paper, we address two issues related to primordial disk evolution in three 
clusters (NGC 1333, IC 348, and Orion A) observed by the INfrared Spectra of Young 
Nebulous Clusters (IN-SYNC) project. First, in each cluster, averaged over the spread 
of age, we investigate how disk lifetime is dependent on stellar 
mass. The general relation in IC 348 and Orion A is that primordial disks around intermediate mass stars 
(2--5$M_{\odot}$) evolve faster than those around loss mass stars (0.1--1$M_{\odot}$),
which is consistent with previous results. 
However, considering only low mass stars, we do not find a significant dependence of disk 
frequency on stellar mass. These results can help to better constrain theories on gas 
giant planet formation timescales. Secondly, in the Orion A molecular cloud, in the mass range
of 0.35--0.7$M_{\odot}$, we provide the most robust evidence to date
for disk evolution within a single cluster exhibiting modest age spread. 
By using surface gravity as an age indicator and employing \mbox{4.5 $\micron$}
excess as a primordial disk diagnostic, we observe a trend of decreasing disk frequency 
for older stars. {The detection of intra-cluster disk evolution in NGC 1333 and IC 348 is tentative, since
the slight decrease of disk frequency for older stars is a less than 1-$\sigma$ effect.}
\end{abstract}



\section{Introduction}
Circumstellar disks are a natural consequence of the conservation of angular momentum during the 
collapse of star forming molecular clouds \citep{2011ARA&A..49...67W}. Excesses above the 
stellar photosphere at short infrared (IR) wavelengths (2--8 $\micron$) trace dust in 
the inner disk \citep{1995ApJ...452..736H}. The excesses generally decrease and finally disappear 
as the disks evolve. 
The disappearance of gas requires accretion into the star, accretion into giant 
planets, or photoevaporation by the radiation from the central star 
\citep{2009apsf.book.....H}. 
The timescale of gas dispersal has been found to be similar to that of dust from measurements
of accretion indicators \citep{Fedele2010}.
This timescale around typical young stars is found to be \mbox{$\sim$3 Myr} \citep{2001ApJ...553L.153H, 
2007ApJ...662.1067H, 2009AIPC.1158....3M}. 

There is clear evidence that disk evolution is dependent on stellar mass. Previous studies 
show that gas-rich disk frequencies around solar to higher mass stars are lower than those
around low mass stars \citep{Hillenbrand1998, 2005AJ....129..856H, 
2006ApJ...651L..49C, 2014MNRAS.442.2543Y, 2014A&A...561A..54R, 2015A&A...576A..52R}, which 
indicates that disks around early-type stars evolve more quickly due to more efficient disk 
dispersal \citep{2009ApJ...695.1210K}. \citet[hereafter Lada06]{2006AJ....131.1574L} 
suggested a maximum disk frequency around stars with spectral type K6--M2 in the partially 
embedded cluster IC 348. \citet{2007AJ....133.2072D} and \citet{2007ApJ...671.1784H} found 
similar peaks in NGC 2362 and the Orion OB1 association, respectively. However, the 
observed decline of disk frequency for the very lowest mass stars was not conclusive due 
to: (1) poisson error; (2) observational bias: at a fixed wavelength (temperature), the 
smaller solid angle of disks around low-mass stars makes it more difficult to distinguish 
disk emission relative to stellar emission; (3) inner disk hole size and system 
inclination: high inclinations or large holes will make it harder to detect disk excesses, 
especially for low-mass stars \citep{Hillenbrand1998}.

The lifetime of gas-rich disks sets a limit to the timescale available for gas giant 
planet formation (from substantial amounts of gas accreted onto the planetary core) 
\citep{2007prpl.conf..573M, 2009ApJ...698....1C}. Given the dependence of disk fraction on 
stellar mass, less time is available for gas giant planet formation around more massive 
stars. However, a competing effect is that disk mass also increases with stellar mass 
\citep{Andrews2013}, so processes dependent on disk mass surface density and 
orbital timescale may proceed faster around stars of higher mass 
\citep{2009IAUS..258..111M}. Observations have shown hints of a higher giant planet 
occurrence rate around more massive stars \citep{2010PASP..122..905J, 
2016PASP..128j2001B}. Hence, examining trends in disk lifetime versus stellar age as a 
function of stellar mass will put a major constraint on theories of planet formation. 

If the age spread of young stellar objects (YSOs) in a young cluster is small enough that 
these YSOs can be assumed to be coeval, and if the spectral types of them are 
known, one can separate the effects of intrinsic color of the photosphere, reddening, and 
intrinsic color excess due to the presence of disk \citep{1997AJ....114..288M}. Then disk 
frequency can be calculated in an unbiased way for an extinction-limited sample as the 
fraction of stars with excess at a certain wavelength as a function of spectral type 
(stellar mass).

Thanks to the INfrared Spectra of Young Nebulous Clusters (IN-SYNC) program, more than 3000 stars 
in NGC 1333, IC 348, and the Orion A molecular cloud have been observed with 
the Apache Point Observatory Galactic Evolution 
Experiment (APOGEE) project from the third Sloan Digital Sky Survey \citep[SDSS-III]{Eisenstein2011}. 
These stars have effective temperatures ($T_{\rm eff}$) and surface gravities (log$g$) determined with a consistent
modeling approach using APOGEE's high resolution $H$ band spectra \citep{Cottaar2014}, and therefore provide 
good samples without any systematic discrepancies in terms of stellar parameters. In this 
paper, using $\sim$2000 of these stars, we address two issues in each of the three clusters.
On one hand, by taking a snapshot of each cluster averaged over the spread of age, we 
investigate how disk frequency is dependent on stellar mass, with particular focus on 
the low-mass range (0.1--1.5 $M_{\odot}$). On the other hand, since 
\citet{Cottaar2014} and \citet{DaRio2016} have detected the intrinsic 
age spreads on the order of $\sim$Myr in these clusters, we are motivated to study disk 
evolution within every single cluster by using log$g$ as an age indicator. This is a novel 
approach to disk evolution problems without having to assume that each cluster is somehow 
representative of a global population.

This paper is organized as follows. The next section provides an assessment of the 
representativeness of our sample. We outline our disk diagnostic in Section 
\ref{sec:diagnostic}. Evidence of disk evolution related to stellar mass and stellar age 
are illustrated in Section \ref{sec:mass} and Section \ref{sec:age}, respectively. A 
conclusion is given in Section \ref{sec:conclusion}. The sample used in this paper can be 
found in Appendix.

\section{Sample Properties} \label{sec:represent}
In this section, we assess the degree to which our sample of stars is considered representative of 
the entire population of cluster members using color magnitude diagram (CMD) methods. First, 
in Section \ref{subsec:21}, we briefly describe the IN-SYNC survey, and the properties of
each cluster being observed. Then, in Section \ref{subsec:22},
we make some initial cuts to the observed IN-SYNC sample to choose a subset of stars for data 
analysis. After that, we individually determine the representative mass 
range under an extinction limit in each cluster in Section \ref{subsec:23}. 
Last, in Section \ref{subsec:caveat}, we point 
out some caveats in our assessment.

\subsection{The IN-SYNC Survey}\label{subsec:21}
The star forming regions targeted by the IN-SYNC program are IC 348 
and NGC 1333 in the Perseus molecular cloud, the Orion A molecular cloud, and NGC 2264. 
Since a relatively small number of sources were observed in NGC 2264, we do not
include this cluster in our study. The APOGEE spectrograph covers the spectral range 
from 1.51 to 1.70 $\micron$ with a spectral resolution of $\sim$22,500. It can also observe up to 
300 targets in a 3 degree diameter field-of-view (FoV) at the same time \citep{Majewski2017}. 
However, it can not simultaneously observe stars within 71.5$''$ of each 
other due to fiber collision. Therefore, multiple plates were drilled to cover the densest regions. 

Target selection for spectroscopic survey of the Perseus fields was described in detail in 
\citet{2015ApJ...799..136F} and \citet{2015ApJ...807...27C}. In summary, potential targets 
were compiled from candidate or confirmed members selected via signatures of youth. 
Observations were 
designed to maximize the completeness for sources with 8 $< H <$ 12.5. For these bright 
sources, high priority was further sorted according to their extinction-corrected $H$ band 
magnitudes. 

The Orion A molecular cloud is the nearest known massive stellar nursery. The large 
structure includes the Upper Sword, NGC 1977, Orion Molecular Cloud 2/3 region (OMC 2/3), 
Orion Nebula Cluster (ONC), $\iota$ Ori (also called NGC 1980), and the low density L1641 
region \citep{Muench2008}. Targets for the IN-SYNC Orion survey have been 
primarily selected from known or candidate members in ONC or L1641, accompanied with 
bright 2MASS sources with unknown membership. Unlike the Perseus survey, observations were 
only conducted for sources with $H <$ 12.5. Details of the observing strategy are 
described by \citet{DaRio2016}. 

\begin{table*}[htb!]
\centering
\caption{Adopted values for each cluster} \label{tab:info}
\begin{tabular}{c|c|c|c|c|cc}
\hline
\hline
Cluster & Mean Age 	& Distance  & Representative   		& Extinction Limit&\multicolumn{2}{c}{SpType}	\\
	    & (Myr) 		& (pc)	   & Mass Range ($M_{\odot}$)	& (mag)			&2.2$M_{\odot}$	&5$M_{\odot}$	\\
\hline
\multirow{3}{*}{NGC 1333} & \multirow{3}{*}{1--2 (a)}	& 282.3    & 0.1--1.5 		&\multirow{3}{*}{$A_{J}<10$} 	& \multirow{3}{*}{K3}	&\multirow{3}{*}{B5}\\
					&					& 300.7	& 0.1--1.5		&						&				&\\
					&					& 329.3	& 0.1--1.5		&						&				&\\
\hline
\multirow{3}{*}{IC 348}       & \multirow{3}{*}{3 (b)}	& 304.2   	&0.23--1.5  &\multirow{3}{*}{$A_{J}<6$}	& \multirow{3}{*}{K1} &\multirow{3}{*}{B5}\\
					&					& 324.2	&0.25--1.5	  &						&				&			\\
					&					& 343.0	&0.27--1.5  &						&				&		\\
\hline
\multirow{3}{*}{Orion A}	 & \multirow{3}{*}{2 (c)}	& 380.0   	& 0.31--1.5	& & \multirow{3}{*}{K2}   &\multirow{3}{*}{B5} \\ 
					&					& 396.7	& 0.34--1.5	&$A_{J}<3$		&			&	\\
					&					& 415.1	& 0.36--1.5	&				&			&\\
\hline
\end{tabular}
\tablecomments{Age references: (a) \citet{Lada1996} (b) \citet{Muench2007} (c) \citet{Muench2008}.
Distances are the 25th, 50th and 75th percentiles of the $Gaia$ DR2 distance distribution of cluster members.
The representative mass range is given at a certain extinction limit.}
\end{table*}

To assess the representative mass range, we need to know the age and distance of each cluster.
The adopted values are presented in Table \ref{tab:info}.
For each cluster, we use the most commonly cited age estimates for consistency 
with other works concerning disk evolution. Distance estimates are derived by 
cross-matching cluster members (compiled in Section \ref{subsubsec:membership})
with the $Gaia$ second data release \citep[hereafter $Gaia$ DR2]{2018arXiv180409365G}.
In each cluster, we calculate the 25th, 50th and 75th percentiles of the distance distribution, as are shown 
in the third column of Table \ref{tab:info}. The median distance in each cluster will be
adopted for assessing the representative mass range.
However, we also provide results corresponding to a tolerant distance estimation (the 25th percentile)
and a restrictive distance estimation (the 75th percentile). See Section \ref{subsec:23} for details.

\subsection{Sample Selection}\label{subsec:22}
Here we present the procedures to select a sample of stars for data analysis 
from the observed targets.

\subsubsection{Membership} \label{subsubsec:membership}
To identify cluster members from the observed sources, we first 
use previous membership studies to select YSO candidates. Then, 
we make restrictions on their $Gaia$ DR2 distances and proper motions to
reduce possible contamination from field stars or background giants.

\paragraph{Previous Studies}
Since target selection for the IN-SYNC survey was performed about five years ago, 
some recently-identified cluster members were not included in the initial catalogue. 
Since un-filled APOGEE fibers were assigned to other
sources within the same filed, however, some were serendipitously observed. 
On the other hand, some stars in the input catalogue 
may display spectral signatures of youth, but are recently rejected as non-members
by proper motion measurements. Therefore, in Perseus, to select cluster members 
from the observed stars, we first combine our input catalogue with the updated census
of NGC 1333 and IC 348 presented by \citet[Table 1 and Table 2]{Luhman2016}, 
and exclude those with proper motions that differ by $>$ 6 mas yr$^{-1}$ from the cluster median (Table
3 of \citet{Luhman2016}). After that, we cross match the compiled
catalogue with the IN-SYNC targets to identify bona fide members observed by IN-SYNC. This 
identifies 104 members in NGC 1333 and 372 in IC 348.

\begin{figure}[htbp]
\includegraphics[width=\columnwidth]{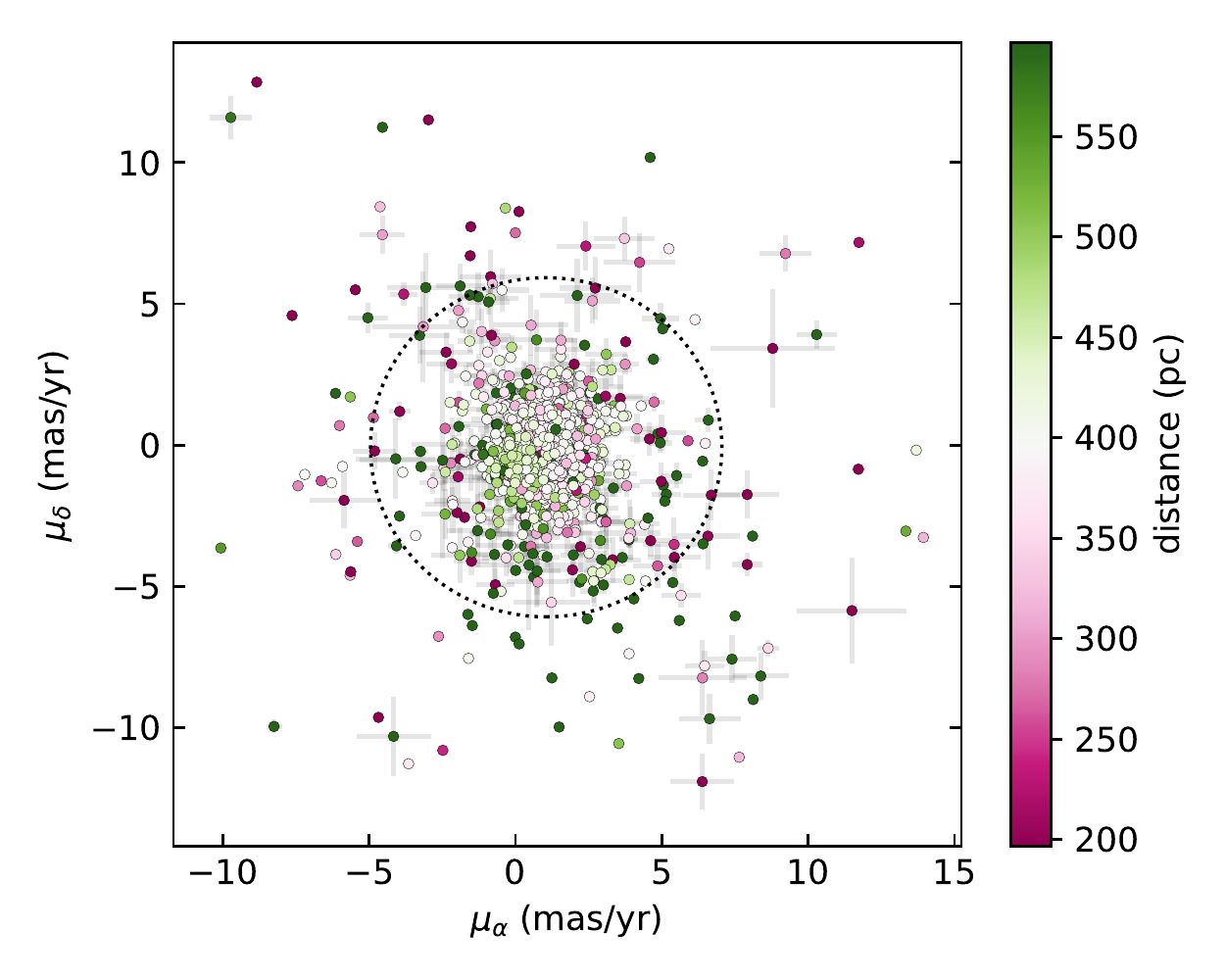}
\caption{Proper motions of 1900 member candidates in Orion A,
color-coded according to distances. 
The dotted line marks a circle with a radius of
6 mas yr$^{-1}$ from the proper motion median ($\mu_{\alpha}=1.05$,
$\mu_{\delta}=-0.08$).
\label{fig:pp_OrionA}}
\end{figure}

\begin{figure*}[htbp]
\centering
\includegraphics[width=0.8\textwidth]{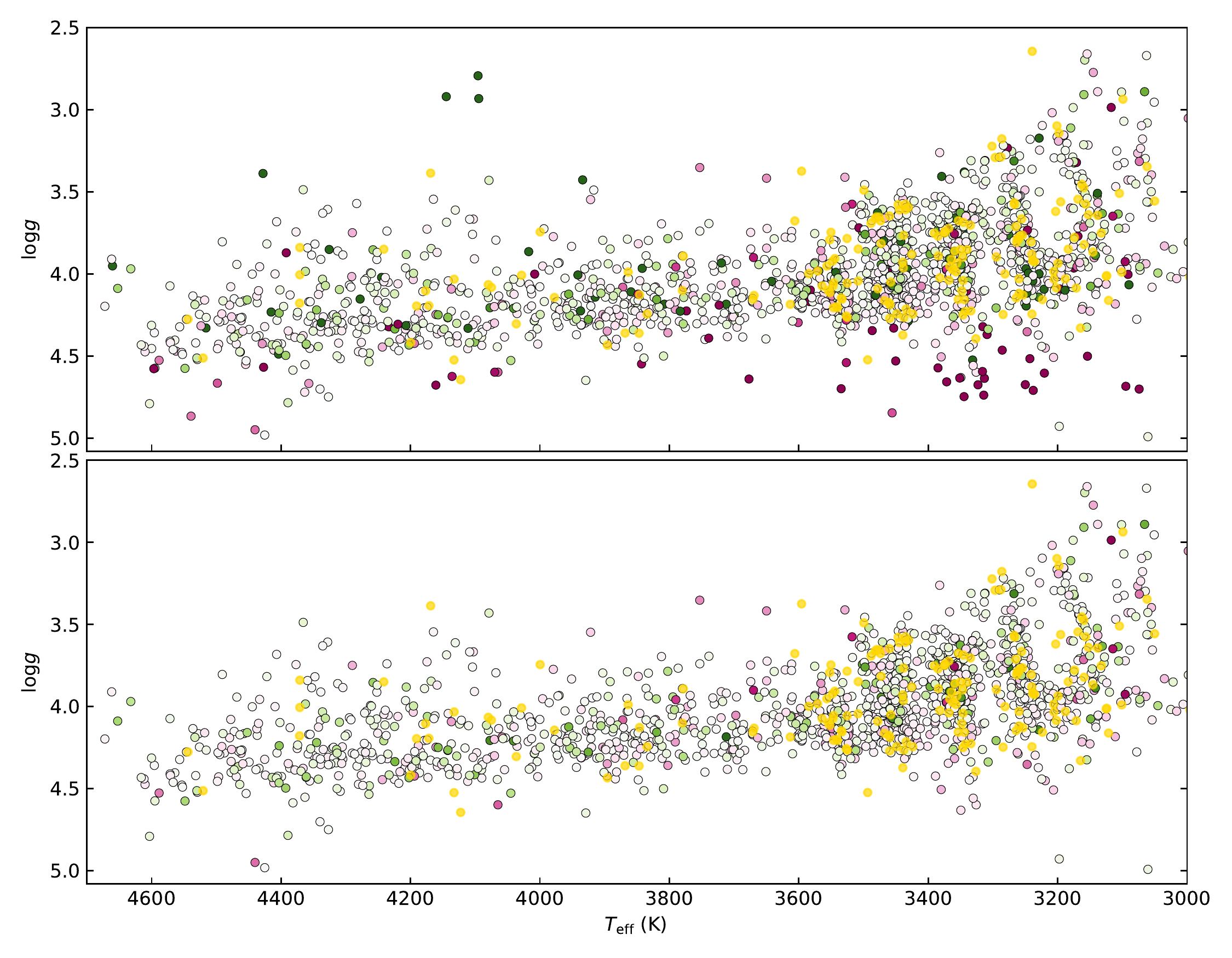}
\caption{Upper panel: distribution of 2092 member candidates in Orion A
on the log$g$--$T_{\rm eff}$ plane.
1900 targets with $Gaia$ DR2 data are color coded by distances (the color scale is the 
same as in Figure \ref{fig:pp_OrionA}). 192 targets without $Gaia$ data are shown
as the yellow dots. Lower panel: distribution of 1837 accepted cluster members, 
including 1645 targets with $Gaia$ data and 192 targets not in the $Gaia$ catalog.
\label{fig:pphrd_OrionA}}
\end{figure*}

Membership identification for Orion is different.
In total, 2691 stars were observed in the Orion A region; 1709 are previously known  
members confirmed by spectroscopy, X-ray emission, or infrared excess; among the remaining 
982 sources, a previous work in the IN-SYNC program \citep{DaRio2016} determines 
whether they are member candidates or not. The authors place all the sources in a number of planes 
(including the CMD $H$--($J-H$) plane, the H--R diagram, the log$g$--$T_{\rm eff}$ plane, the position--velocity plane), 
and compare the location occupied by each of them with the region occupied by the 1709 known 
members, since the latter generally occupy well-confined regions. 
For each star with no membership, in each plane, \citet{DaRio2016} consider the 
50 closest sources to the star, 
and count among the 50 sources the fraction that are previously known members. 
This value is a basic indicator of its membership probability. The method identified 
383 new candidate members, with 599 ($=982-383$) stars considered to be non-members.
We remove these 599 stars and leave 2092 in the Orion A sample.

\paragraph{$Gaia$ Distances and Proper Motions}
Section 5 of \citet{DaRio2016} provides a discussion of the limitation of their method.
In brief, newly-identified members are only candidates, but not confirmed members.
Possible contamination can be further eliminated if distances and proper motions are know.
Therefore, we cross-match the 2092 sources in Orion A with the $Gaia$ DR2 catalog;
1900 of them have the ``astrometric global iterative solution" (AGIS) including 
position, parallax, and proper motion. Their proper motions are shown in Figure \ref{fig:pp_OrionA},
color-coded according to distances. The colormap is adjusted such that the deeper the color,
the further the star is from the median distance of 396.7 pc (Table \ref{tab:info}).
We reject objects either with motions that differ by more 
than 6 mas yr$^{-1}$ from the median of proper motion (beyond the dotted
circle in Figure \ref{fig:pp_OrionA}), 
or with distances that
differ by more than 200 pc from the 396.7 pc. The choice of 
6 mas yr$^{-1}$ and 200 pc is based on inspection of Figure \ref{fig:pp_OrionA}.
1645 of the 1900 candidates survive these cuts. 

We demonstrate the importance of this step in Figure \ref{fig:pphrd_OrionA}.
Distribution of all of the 2092 member candidates on the log$g$--$T_{\rm eff}$ plane
is shown in the upper panel. 1900 targets with $Gaia$ DR2 data are color-coded by
distances, while the 192 sources without $Gaia$ data are shown as yellow dots. 
In Section \ref{subsubsec:stellar parameters}, we will only select targets
with \mbox{3000 K $ <T_{\rm eff}<4750$ K}. Hence, only sources in this temperature range are displayed.
At the bottom right of this panel, there is clearly a cluster of deep purple dots
with very high log$g$ and small distances. They are probably contaminants of field stars.
There are also a few distant stars locating around \mbox{$T_{\rm eff} \approx4300$ K}, log$g\approx 2.9$,
which should be contaminants of background red giant stars.  

The lower panel of Figure \ref{fig:pphrd_OrionA} shows distribution of the 1645
targets surviving cuts on proper motion and distance, as well as the 192 sources without
$Gaia$ data. Most contaminations from both field stars and background giants are excluded.
Although we are unable to make cuts on the 192 yellow dots, they
generally lie along the majority of cluster members on the log$g$--$T_{\rm eff}$
diagram, and may contain very few contaminants. Therefore, we choose not to exclude
any sources from them, and the number of sources retained in Orion A is reduced to 1837 $(=1645+192)$.

To make sure that our analysis is consistent in all of the three clusters,
we also apply the procedures stated above in the other two clusters. After that,
the number of sources decreases from 104 to 95 in NGC 1333,
and from 372 to 324 in IC 348. 

\subsubsection{Stellar Parameters} \label{subsubsec:stellar parameters}
\begin{figure}[ht!]
\centering
\includegraphics[width=0.9\columnwidth]{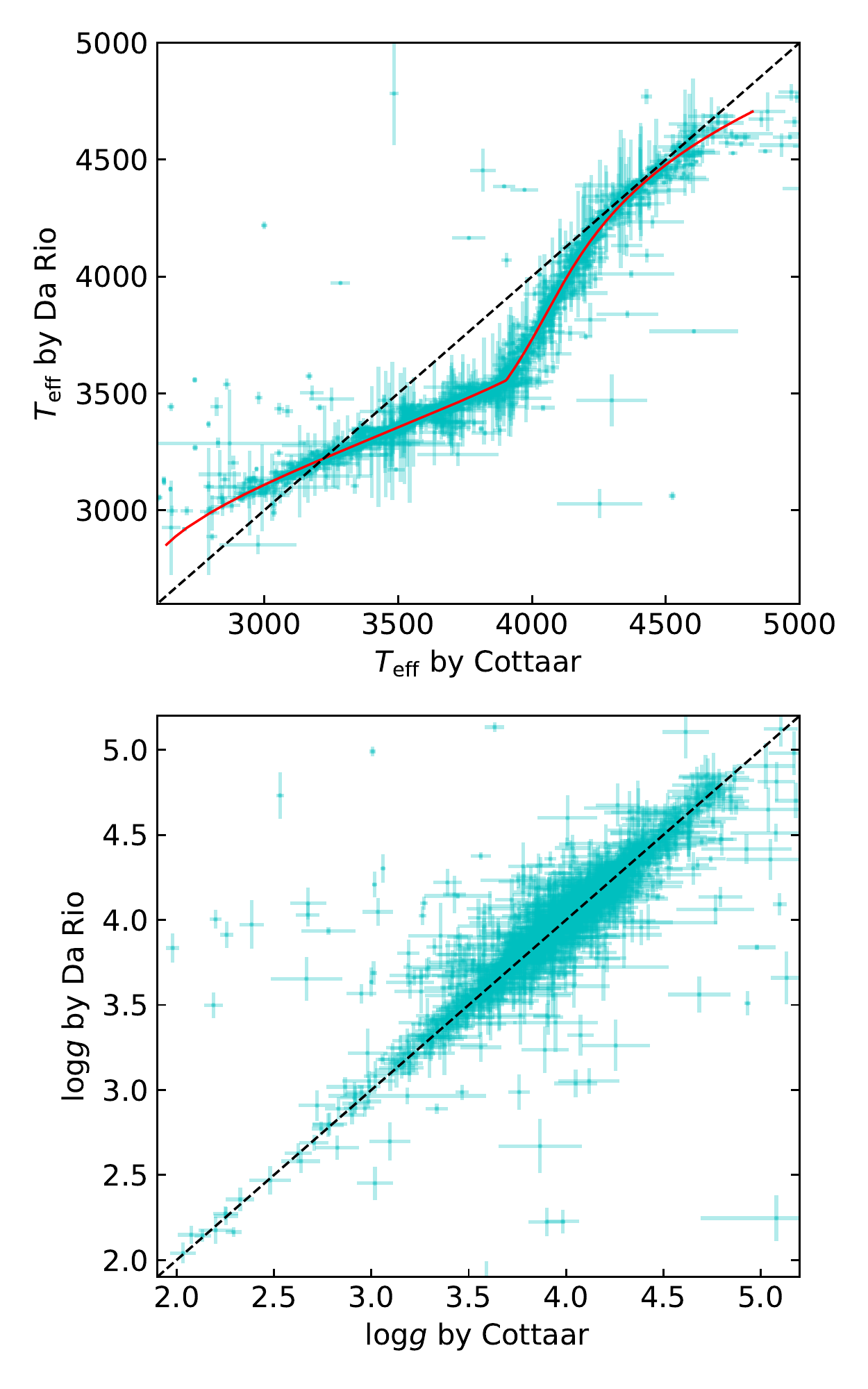}
\caption{Comparison of $T_{\rm eff}$ and log$g$ determined for IN-SYNC Orion A targets by 
\citet{Cottaar2014} and \citet{DaRio2016}. The red line in the upper panel is a spline fit of data points,
while the dashed black lines in both panels are just diagonal lines of $y=x$.
\label{fig:comparePars}}
\end{figure}

All IN-SYNC spectra were modeled by a spectral fitting approach presented in \citet{Cottaar2014},
which is suited for determining stellar parameters of young stars. 
In brief, observed spectra are modeled with a grid of ``BT-Settl'' synthetic spectra \citep{Allard2012}.
Five free parameters are included in the fitting: $T_{\rm eff}$, 
log$g$, radial velocity ($v_{r}$), rotational velocity ($v$sin$i$), and $H$-band veiling 
($r_{H}$). For targets with multiple observations, we adopt the average values 
determined as the weighted mean 
of the parameters measured from all epochs.
Parameter uncertainties are initially estimated from a 
Markov Chain Monte Carlo (MCMC) simulation, and then 
inflated to match the epoch-to-epoch variability seen 
for the same star at different epochs.

Although spectra in all clusters were initially modeled by \citet{Cottaar2014},
\citet{DaRio2016} implemented changes to correct and optimize aspects of the code,
and utilized the updated algorithm to model spectra of sources in Orion A. Results from 
the latter paper should supersede those from \citet{Cottaar2014}.
Unfortunately, we do not have the new parameters for sources in Perseus, so we 
still need to use Cottaar's parameters for NGC 1333 and IC 348. In Figure \ref{fig:comparePars},
we compare the $T_{\rm eff}$ and log$g$ determined by the two authors for sources in Orion A. 
In the upper panel, it is obvious that there is a systematic discrepancy of $T_{\rm eff}$
determined by the two methods.
This offset is most prominent near \mbox{$T_{\rm eff}\approx3900$ K}, where the Da Rio values are several hundred kelvins
cooler than the Cottaar results. As \citet{DaRio2016} found no such offset in their comparison to
$T_{\rm eff}$ estimates in the literature (see their Figure 3), we adopt the scale implied by the Da Rio results.
To this end, we fit a spline function of $T_{\rm eff}$ (Cottaar) vs. $T_{\rm eff}$ (Da Rio) to the data points, 
which is shown as the red line. Cottaar's temperatures of the Perseus sources are then scaled 
with this relation to bring them into agreement with Da Rio's. 
By inspection of the lower panel, we see no discrepancies for the surface gravity measurements.
The data generally follow a linear relation of log$g$ (Cottaar) $=$ log$g$ (Da Rio). 
Therefore, for sources in Perseus, we do not make any scaling to log$g$ produced by Cottaar.

\citet{Cottaar2014} and \citet{DaRio2016} also find that APOGEE-$T_{\rm eff}>$ 4750K is less reliable 
due to the smaller number of features in $H$-band spectra for hotter stars.
Therefore, we only retain stars with 3000 K $<T_{\rm eff}<4750$ K.
This subset of cluster members should have accurately determined $T_{\rm eff}$ for data analysis.
After this temperature cut, the number of stars decreases from 95 to 75 in NGC 1333, 
from 324 to 263 in IC 348, and from 1837 to 1702 in Orion A.
The average age of each cluster ranges from 1 to 3 Myr (Table \ref{tab:info}). According to the \citet[hereafter 
BCAH98]{1998A&A...337..403B} pre-main-sequence (PMS) evolutionary models, for stars with 
ages between 1 and 3 Myr, $T_{\rm eff}=3000$ K (4750 K) corresponds to $\sim$0.1 $M_{\odot}$ 
(1.5 $M_{\odot}$) stars. 

\subsubsection{Photometry} \label{subsubsec:photometry}
\paragraph{IRAC}
We utilize infrared photometry from online archives of the $Spitzer$ 
Infrared Array Camera (IRAC) instrument, which includes four bands at 3.6, 4.5, 5.8, and 8.0 $\micron$. 
To obtain an idea of IRAC's detectability, we calculate the expected photospheric IRAC flux
of a $T_{\rm eff}=$ 3000 K, log$g=$ 4.0 star by convolving synthetic ``BT-Settl'' 
photospheres \citep{Allard2012} with the response curve of IRAC filters. 
The choice of $T_{\rm eff}$ and log$g$ is typical of a low mass ($\sim$0.1$ M_{\odot}$) young star.
Assuming a distance of 420 pc and an extinction of $A_{J}=6$, this star would appear to be a source
with $I_1=13.062$, $I_2=12.696$, $I_3=12.586$, $I_4=12.578$, well above the limit magnitude of
IRAC. Since all sources in our sample are hotter (3000 K $<T_{\rm eff}<4750$ K), 
closer ($\sim$280--415 pc), and suffer from less extinction, they should be detected by
IRAC even if there is no disk emission.

IRAC data for the Perseus clusters are obtained from the c2d 
$Spitzer$ Legacy Project \citep{2003PASP..115..965E, 
2009ApJS..181..321E}\footnote{http://irsa.ipac.caltech.edu/data/SPITZER/C2D/}. 
Among the 75 stars in NGC 1333, 1 star (2MASS J03290915+3121445, $H=14.8$) 
has no data in the c2d catalog. 
The online image shows that it is only about $20''$ from the center of the nebula. 
It is suspected that the non-detection may arise from a technical issue.
We remove this star, and retain 74 for further analysis. 
Among the 263 stars in IC 348, 17 are outside the FoV of the c2d survey.
We retain the 246 sources within the FoV of the IRAC camera.

IRAC data for Orion A are obtained from the Spitzer Enhanced Imaging Products (SEIP) 
source list\footnote{http://irsa.ipac.caltech.edu/data/SPITZER/Enhanced/SEIP/}. To ensure 
high reliability, SEIP remove many observed sources by strict cuts in size, blending, 
signal-to-noise ratio, etc. Hence, in the most crowded region of ONC, the list is highly incomplete. We use 
the 3.8$''$ diameter aperture flux density (including band-filled fluxes when the IRAC 
source is extended). 1431 sources have extracted SEIP fluxes at 4.5 $\micron$. We note that 
although IRAC data for another $\sim$100 stars can be obtained from \citet[the $Spitzer$ 
Orion survey]{2012AJ....144..192M}, we only retain the 1431 SEIP sources, because 
\citet{2012AJ....144..192M} only published sources with infrared excess, which means most 
(if not all) of these $\sim$100 sources have disks. Thus, including them will bias our 
sample to stars with disks and render disk frequency higher than the true value. 

\paragraph{2MASS}
 Three sources in IC 348 fail to be cross-matched with the Two Micron All Sky Survey 
(2MASS) all-sky point source catalog \citep{2006AJ....131.1163S}. 
They are removed since 2MASS colors are needed for our estimates of extinction 
and color excess (see Section \ref{sec:diagnostic}). The number of sources in the IC 348 sample
decreases from 246 to 243.

\subsection{Representative Mass Range}\label{subsec:23}

Bearing in mind that our sample is far from being complete at the low mass end, in each 
cluster we study the range of stellar mass that is considered to be representative by our 
sample. Isochrones used in this section are calculated by BCAH98 with 
a convection mixing length of $\alpha=1.9$. 

\subsubsection{NGC 1333}\label{subsubsec:represent NGC 1333}
\begin{figure}[htbp]
\includegraphics[width=\columnwidth]{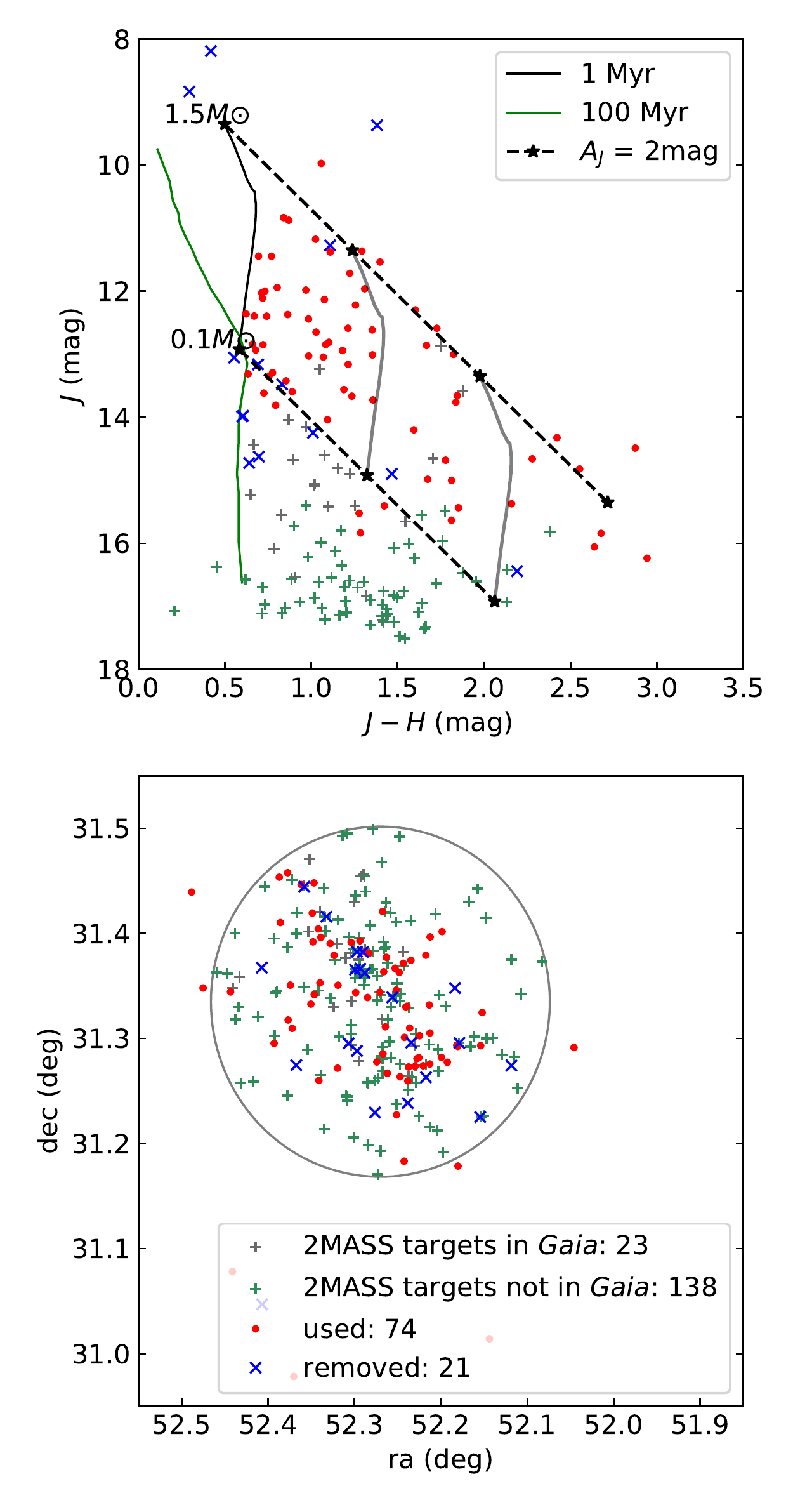}
\caption{Upper panel: Color-Magnitude Diagram (CMD) of sources in NGC 1333. Over-plotted 
isochrones are from BCAH98. 74 observed cluster members in our sample are shown as red dots, 21
observed members not in our sample are shown as blue crosses. Other 2MASS sources within 
10$'$ from the star 2MASS J03290832+3120203 are marked as pluses. Among them, the 138 sources
without $Gaia$ data are shown in green; the 23 sources with $Gaia$ proper motion and distance
measurements being consistent with being cluster members are shown in grey (see text). An extinction 
vector of $A_{J} =$ 2 mag is the length of the dashed black line between two black asterisks. 
Lower panel: spatial distribution of these sources. \label{fig:cmd_NGC1333}}
\end{figure}

The final sample of 74 sources in NGC 1333 are shown as the red dots
in Figure \ref{fig:cmd_NGC1333}. 
The 21 stars shown with blue crosses are also observed by IN-SYNC, 
but removed from data analysis. The red and blue add up to
the 95 cluster members retained in Section \ref{subsubsec:membership}.
\citet{Pecaut2013} provide intrinsic color ($J-H$) and $T_{\rm eff}$ for $<30$ Myr old 
PMS stars.
We convert $T_{\rm eff}$ to stellar mass and $M_{J}$ using the BCAH98 isochrone of 1 Myr stars.
The black line in the upper panel of Figure \ref{fig:cmd_NGC1333} shows the expected $J$ vs. ($J-H$) isochrone
adopting a median distance of 300.7 pc for this cluster (see Table \ref{tab:info}).

We use the following method to select possible cluster members not targeted by IN-SYNC.
First of all, we collect all 2MASS sources within 10$'$ from the center of NGC 1333.
Most cluster members and few field stars are expected to be inside of this circular field. 
The radius of 10$'$ is chosen based on inspection of the spatial distribution of cluster members.
Among these 2MASS sources, targeted stars (red and blue) and
known field stars tabulated in \citet[Table 3]{Luhman2016} are removed, resulting in 221 sources. 
Then, to further exclude contamination of non-cluster members in the 221 sources, 
we cross-match them with $Gaia$ DR2. For the 83 sources with proper motion and distance measurements, 
we reject stars either with motions that differ by more than 6 mas yr$^{-1}$
from the median proper motion of cluster members ($\mu_{\alpha}=7.48$, $\mu_{\delta}=9.89$,
found in Section \ref{subsubsec:membership}),
or with distances smaller than 100.7 ($=300.7-200$) pc or greater than 500.7 ($=300.7+200$) pc, 
where 300.7 pc is the median distance of cluster members. 
The 23 sources that pass these cuts
are shown as grey pluses in Figure \ref{fig:cmd_NGC1333}.
Note that this criterion is also adopted to reject contaminants in 
Section \ref{subsubsec:membership}. The 138 2MASS sources without $Gaia$
data are shown as green pluses. 

\begin{figure}[ht!]
\centering
\includegraphics[width=0.8\columnwidth]{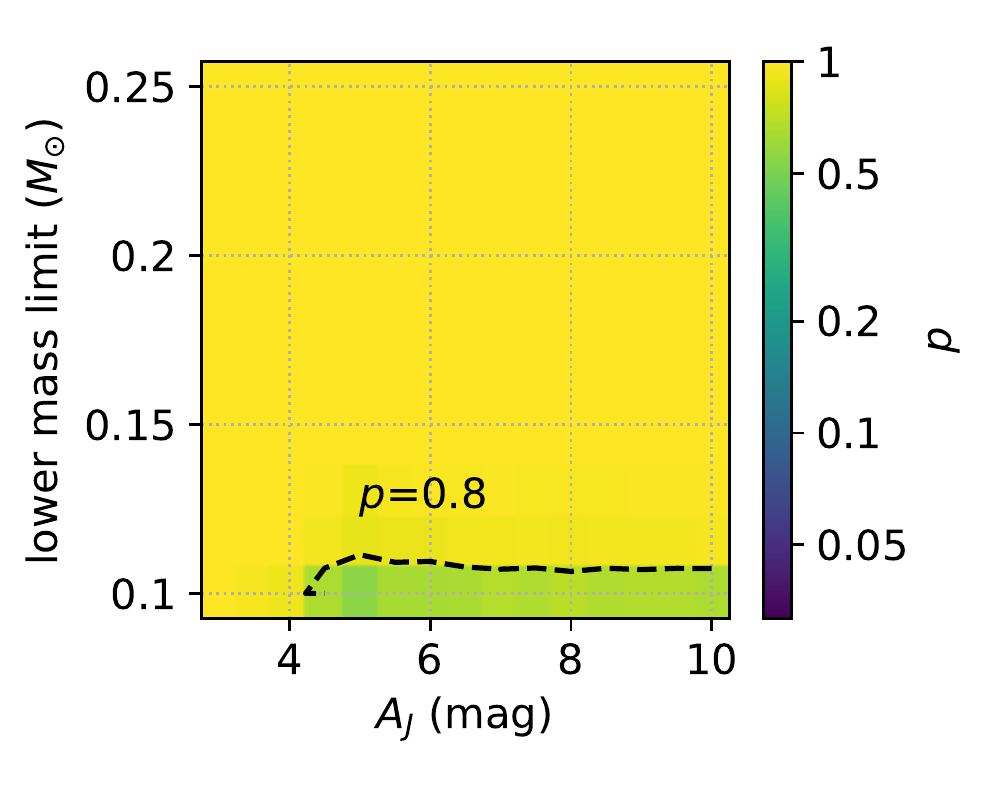}
\caption{K-S test significance level ($p$) for the null hypothesis that
the extinction-limited, mass-limited NGC 1333 sample is uniformly drawn from the total of NGC 1333.
$X$- axis shows the extinction limit, while $Y$-axis shows the lower mass limit.
A contour of $p=0.8$ is shown as the dashed black line. \label{fig:kstest_NGC1333}}
\end{figure}

To assess the representativeness of the used sample in NGC 1333 (red dots in Figure 
\ref{fig:cmd_NGC1333}), we perform a two-sample 2D K-S test to test the null hypothesis 
that the used sample is drawn from the same distribution as that of the total cluster members. 
Here we assume that the total cluster members can be represented by observed cluster
members (red + blue) combined with other 2MASS stars within the adopted cluster radius (grey 
+ green). The applicability of including these 2MASS sources in the K-S tests is 
discussed in Section \ref{subsubsec:use2mass}.

This test is utilized to see if the used sample is uniformly selected from the total, or 
if it represents a subsample that is not consistent with the total. 
It is performed at a certain extinction limit of $A_{J}$, and in a certain mass range.
In short, in the upper panel of Figure \ref{fig:cmd_NGC1333}, in the corresponding 
zone of the CMD, 2D K-S test 
ranges over every point on the ($J-H$, $J$) plane to calculate the fraction of points 
(i.e., probability) from the two samples (red vs. red+blue+grey+green) in each of the four 
natural quadrants, in search of the maximum probability difference (defined as $D$) ranging both over data 
points and over quadrants \citep[Section 14.7]{Press2002}. 

\begin{figure}[ht!]
\includegraphics[width=\columnwidth]{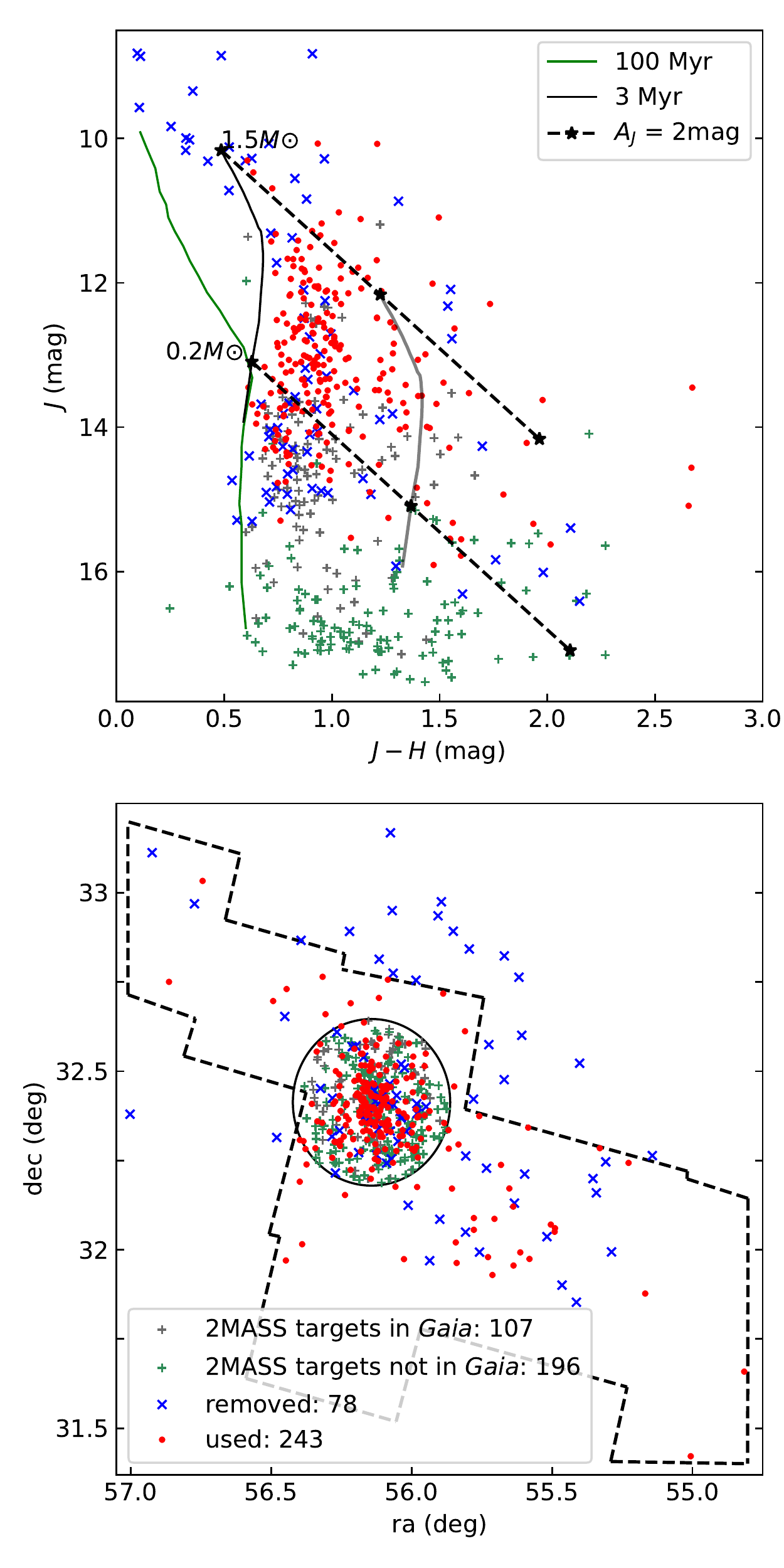}
\caption{Upper panel: CMD of sources in IC 348. 243 observed 
sources in our sample are shown as red dots, 78 observed sources not in our sample are 
shown as blue crosses. Other 2MASS sources within 14$'$ from the star B5 star HD 281159 
are marked as grey and green pluses (see text). An extinction vector of $A_{J} = $ 2 mag is 
shown as the length between two black asterisks on the dashed black line. 
Lower panel: spatial distribution of these sources. The dashed black line 
shows the IRAC field of view. \label{fig:cmd_IC348}}
\end{figure}

To know if $D$ is statistically significant, we generate 1000 synthetic data sets by randomly choosing
some points from the total (allowing repeat). Defining $N_1$ to be the number of points in the sample,
and $N_2$ to be that in the total, each data set has $N_1$ points. 
We then compute $D$ for each synthetic data set, and count what fraction of the 1000 synthetic $D$ exceeds the 
$D$ from the real sample. This fraction is the significance level ($p$-value)\footnote{We have also computed $D$
using Eq. (14.7.1) of \citet{Press2002}, which evaluates if $\frac{\sqrt{N}D}{1+\sqrt{1+r^2}(0.25-0.75/\sqrt{N})}$
is large enough to reject the null hypothesis, where $N = N_1N_2/(N_1+N_2)$ and $r$ is Pearson's 
coefficient of correlation. This equation treats $N_1$ and $N_2$ equally and gives us very similar but slightly higher $p$-value.}.
One might consider a hypothesis suspect of $p<0.05$, but here we adopt a conservative value of 
$p>0.2$ to insist that the null hypothesis is retained.
The upper mass limit is set at 1.5 $M_{\odot}$, and we let the lower mass limit vary from 0.1 to 0.25 $M_{\odot}$.
The extinction limit of $A_J$ changes from 3 to 10, because among all of the 74 stars in the sample,
the highest extinction is $A_J=9.2$ (see Section \ref{subsec:determine excess} for the determination of extinction).
The resulting $p$ is shown as the colormap in Figure \ref{fig:kstest_NGC1333}. 
In the parameter space being investigated,
the returning $p$-values are above 0.2.
We then conclude that our sample is representative 
for 0.1--1.5 $M_{\odot}$ at $A_{J}=10$ in NGC 1333.

\subsubsection{IC 348} \label{subsubsec:represent IC 348}
\begin{figure}[ht!]
\centering
\includegraphics[width=0.9\columnwidth]{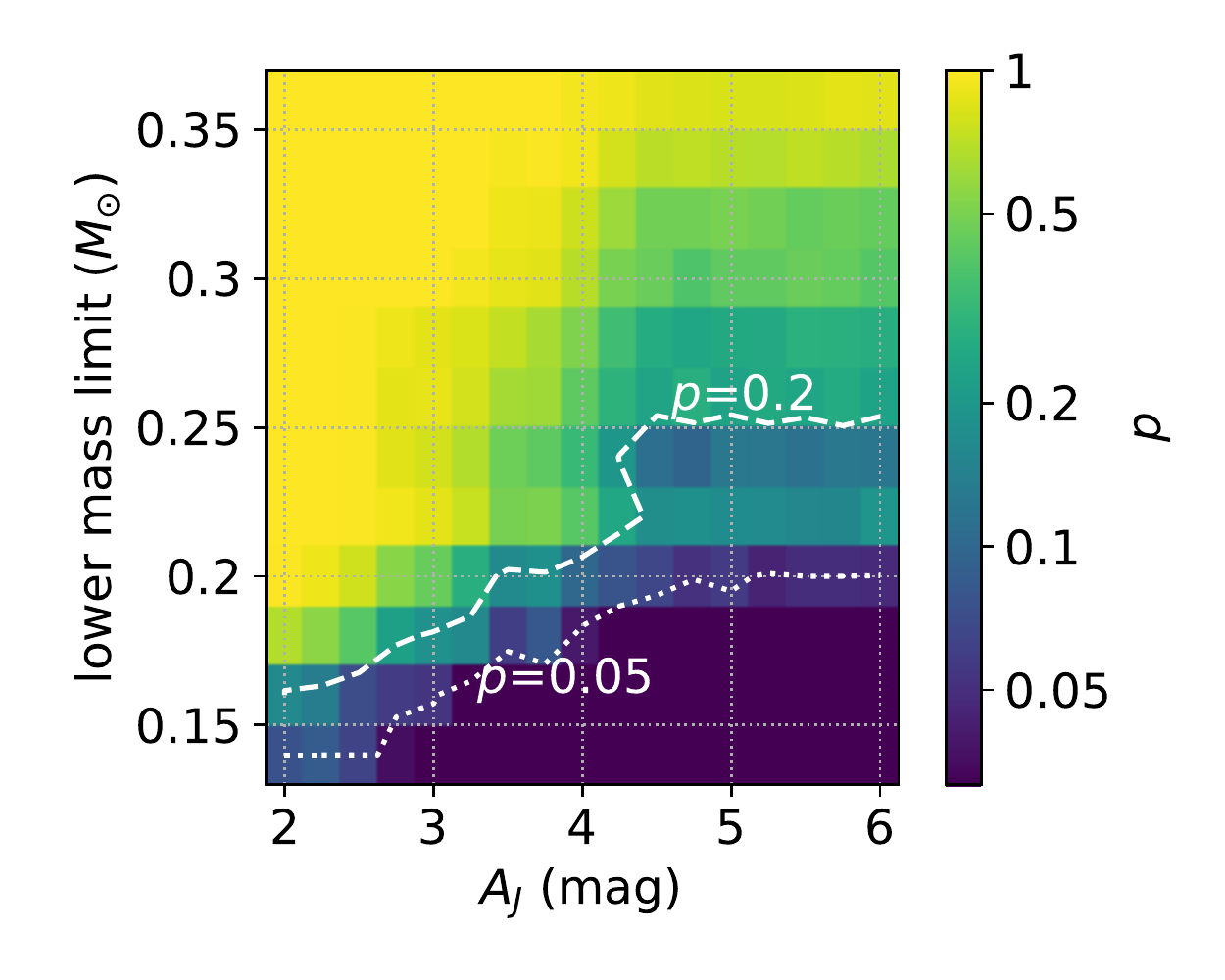}
\caption{K-S test significance level ($p$) for the null hypothesis that
the extinction-limited, mass-limited IC 348 sample is uniformly drawn from the total.
See Figure \ref{fig:kstest_NGC1333}. \label{fig:kstest_IC348}}
\end{figure}

\begin{figure*}[ht!]
\includegraphics[width=\textwidth]{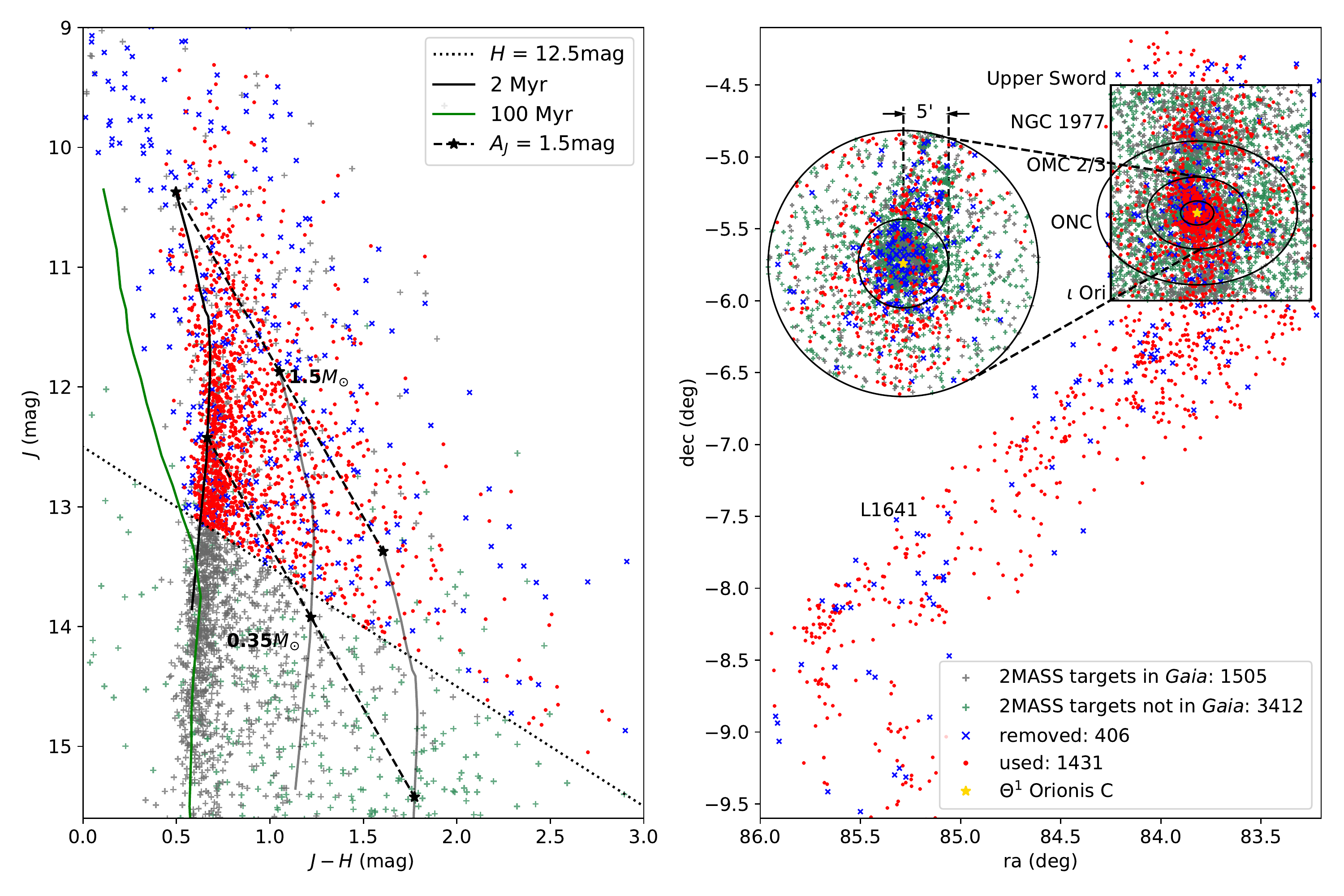}
\caption{Left panel: CMD of sources in Orion A.
1431 observed members in the final sample are shown as red dots, 406 observed members
not in the final sample are shown as blue crosses. 
Other 2MASS sources not observed are marked as grey and green pluses (see text).
An extinction vector of $A_{J}=1.5$ is shown as the length between two asterisks on the dashed black line.
Right panel: spatial distribution of these sources.
The black square marks the field containing the Orion Nebula, NGC 1977, and OMC-2/3 
(following Figure 14 of \citet{2012AJ....144..192M}). The yellow asterisk marks the location
of $\Theta^1$ Orionis C, which is a O6 star at the center of ONC. \label{fig:cmd_ONC}}
\end{figure*}

In Figure \ref{fig:cmd_IC348}, we show the 243 stars in the final sample of IC 348
as the red dots. The 78 blue crosses are also observed cluster members that are removed from 
further analysis. The red and blue add up to 321, which is not the 324 sources retained in 
Section \ref{subsubsec:membership}, because 3 sources do not have 2MASS data.
We convert $T_{\rm eff}$ in \citet{Pecaut2013} to stellar mass and 
$M_{J}$ using the BCAH98 isochrone of 3 Myr stars.
The black line in the upper panel of Figure \ref{fig:cmd_IC348} shows the expected isochrone
adopting a median distance of 324.2 pc for this cluster (see Table \ref{tab:info}).

Similar to what we have done for NGC 1333, 2MASS sources within 14$'$ from the center of IC 348 
are collected. The radius of 14$'$ is also selected based on inspection of the lower panel of
Figure \ref{fig:cmd_IC348} to include most cluster members and few field stars. 
Again, to reduce contamination, we remove known field stars tabulated 
in \citet[Table 3]{Luhman2016}, which gives us 708 sources to be corss-matched with $Gaia$. 
Among the 512 sources with $Gaia$ data,
107 satisfy the cut of proper motion (no more than 6 mas yr$^{-1}$ away from the cluster median) 
and distance (no more than 200 pc away from the cluster median). They are shown as grey pluses in
Figure \ref{fig:cmd_IC348}, and the 196 2MASS sources without $Gaia$ data are marked with 
green pluses.

The resulting $p$-value of the 2D K-S test is shown in Figure \ref{fig:kstest_IC348}.
The extinction limit of $A_J$ is chosen to vary from 2 to 6 since the highest extinction in this cluster
($A_J=5.5$) is no greater than 6. By inspection of Figure \ref{fig:kstest_IC348},
we conclude that our sample is representative for 0.25--1.5 $M_{\odot}$ at $A_{J}=6$ in IC 348.

\subsubsection{Orion A}\label{subsubsec:data OrionA}
In Figure \ref{fig:cmd_ONC}, the 1431 red points are stars in our final
sample of Orion A, the 406 blue pluses are considered to be cluster members, 
but are removed from analysis. Red and blue add up to the 1837 stars 
retained in Section \ref{subsubsec:membership}.
Since the Orion structure extends over a large spatial area, 
it is relatively difficult to selected unobserved cluster members from 2MASS sources.
First of all, including all other 2MASS sources located within the large structure may introduce 
many contaminants in the total. Hence, we only take the 8650 sources in the FoV of the densest 
region ($-6\degree<{\rm dec}<-4.5\degree$, 
$83.25\degree<{\rm ra}<84.25\degree$) into consideration. 
The 599 non-member (candidates) identified by \citet{DaRio2016}
are excluded to remove contaminants.
Among the 8650 sources, 5238 have $Gaia$ data, with 1505 satisfying the astrometry 
criteria for cluster members applied in Section \ref{subsubsec:represent NGC 1333} 
and \ref{subsubsec:represent IC 348}.
They are shown as grey pluses in Figure \ref{fig:cmd_ONC}.
The 3412 sources not in $Gaia$ are shown as green pluses.

\begin{figure}[ht!]
\centering
\includegraphics[width=\columnwidth]{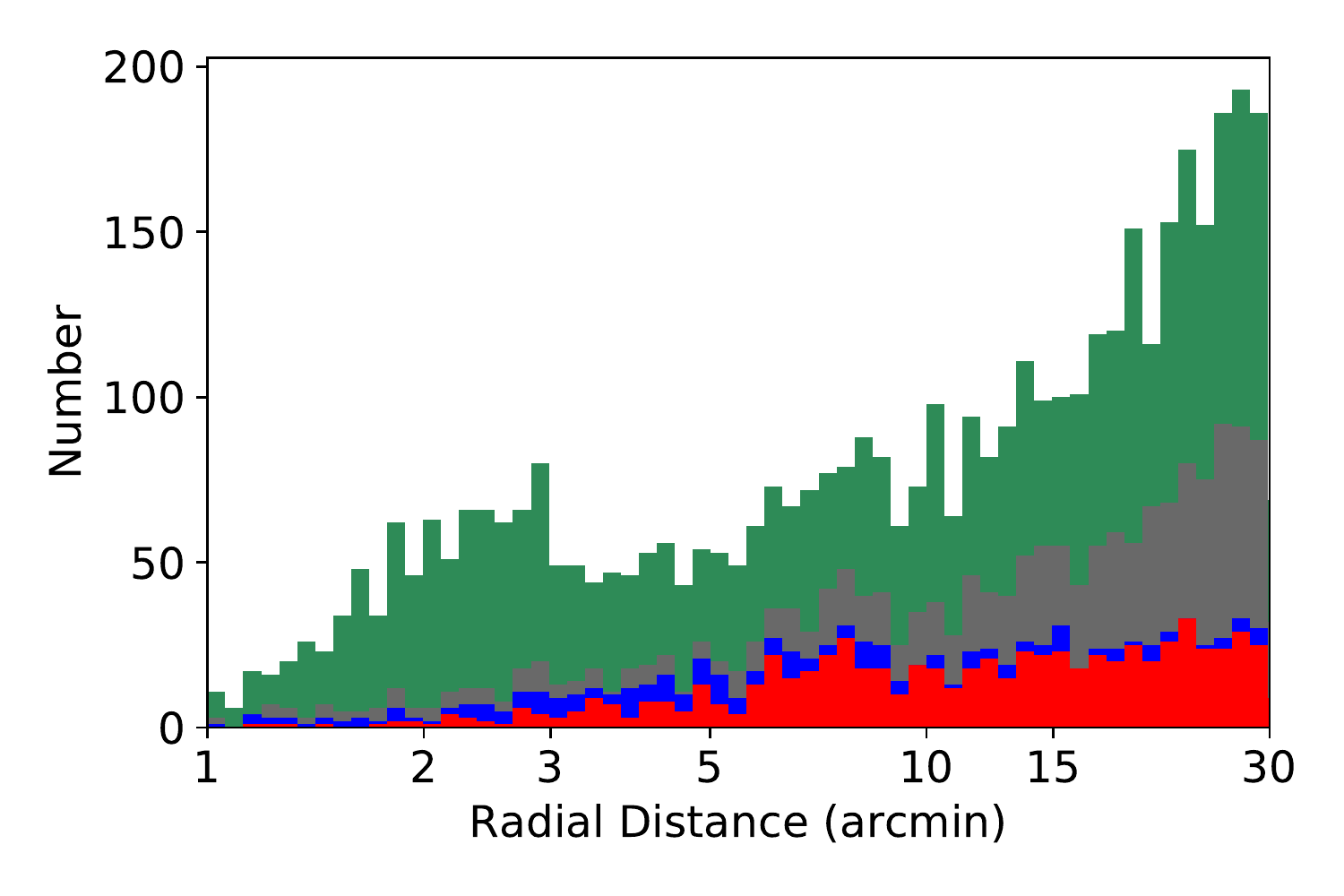}
\caption{Number distribution as a function of radial distance from $\Theta^1$ Orionis C
of cluster members in our sample (red),
cluster members removed from our sample (blue), 2MASS sources
with $Gaia$ data that survive the cuts on proper motion and distance (grey), and 2MASS sources
not in the $Gaia$ catalog (green). 
Only the range of radial distance from 1$'$ to 30$'$ is shown.
The $X$-axis is in the logarithm scale. \label{fig:radial}}
\end{figure}

Another potential caveat is that the representative 
mass range and extinction limit of our Orion A sample may depend on 
radius (the distance between individual stars from the cluster center),
e.g., due to the presence of differential extinction. 
In the right panel of Figure \ref{fig:cmd_ONC}, three circles with radii 
of 5$'$ (0.58 pc), 15$'$ (1.73 pc), and 30$'$ (3.46 pc) from $\Theta^1$ Orionis C 
(marked as the yellow asterisk) are drawn.
We show a zoom-in of the region within 15$'$ from the central star. It can be seen that 
our sample is less complete in the crowded center. 
This inhomogeneity can also be verified in Figure \ref{fig:radial},
where a stacked histogram of sources as a function of radius is shown.
The colors have the same meaning as in Figure \ref{fig:cmd_ONC}.
From a radial distance of 1$'$ to 5$'$, the ratio of the number of stars in our sample (red) to
the total (red+blue+grey+green) is just $\sim$7\%,
while from 5$'$ to 30$'$, this value increases to $\sim$19\%.

\begin{figure}[ht!]
\centering
\includegraphics[width=\columnwidth]{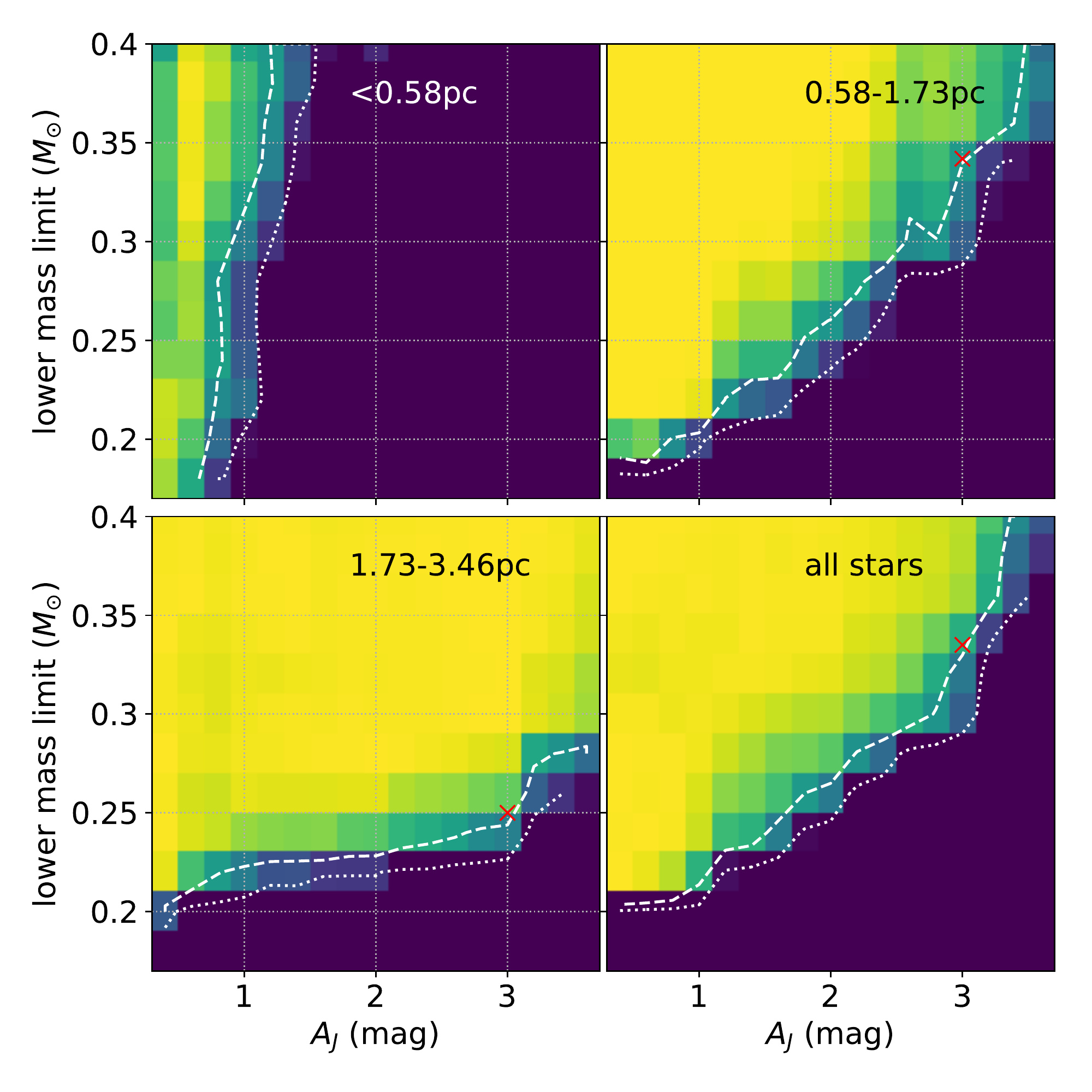}
\caption{K-S test significance level ($p$) for the null hypothesis that
our extinction-limited, mass-limited Orion A sample is uniformly drawn from the total.
The upper left panel considers all sources within 5$'$ (0.58 pc) from $\Theta^1$ Orionis C.
The upper right and lower left panels consider sources from 5$'$ to 15$'$ and
from 15$'$ to 30$'$ from $\Theta^1$ Orionis C, respectively (see the right panel of
Figure \ref{fig:cmd_ONC} for the location of the two annuli). The lower right panel considers all sources 
shown in Figure \ref{fig:cmd_ONC}.  The dashed and dotted lines mark the contours of $p=0.2$
and $p=0.05$, respectively.
The scale of the colorbar is the same as in Figure \ref{fig:kstest_NGC1333} and 
\ref{fig:kstest_IC348}.
\label{fig:kstest_OrionA}}
\end{figure}

\citet{Hillenbrand1998} has found that the
optically-thick disk fraction in ONC increases towards the cluster center.
If the representative mass range of our sample is a function of radius, 
then this could be a systematic effect for the estimated disk frequencies.
Therefore, we assess the representativeness of our Orion A sample
by separating them into different radial bins, and individually applying
the 2D K-S test (assuming an age of 2 Myr and a distance of 396.7 pc). 
The results are presented in Figure \ref{fig:kstest_OrionA}.
In each panel, the red cross lying on the $p=0.2$ contour marks
the lower mass limit at the extinction limit of $A_J=3.0$.
It is evident from this figure that the significance level ($p$) returned by the test increases as we go to outer radius,
indicating that our sample is less representative in the inner part. 
Therefore, we exclude 88 red points within 5$'$ from $\Theta^1$ Orionis C
from our sample, and conclude that the new sample with 1343 ($=1431-88$) YSOs
is representative from 0.34 to 1.5 $M_{\odot}$ at the extinction limit of $A_J=3.0$,
which is the location of the red cross in the upper right panel of Figure \ref{fig:kstest_OrionA}.

\begin{figure}[ht!]
\includegraphics[width=\columnwidth]{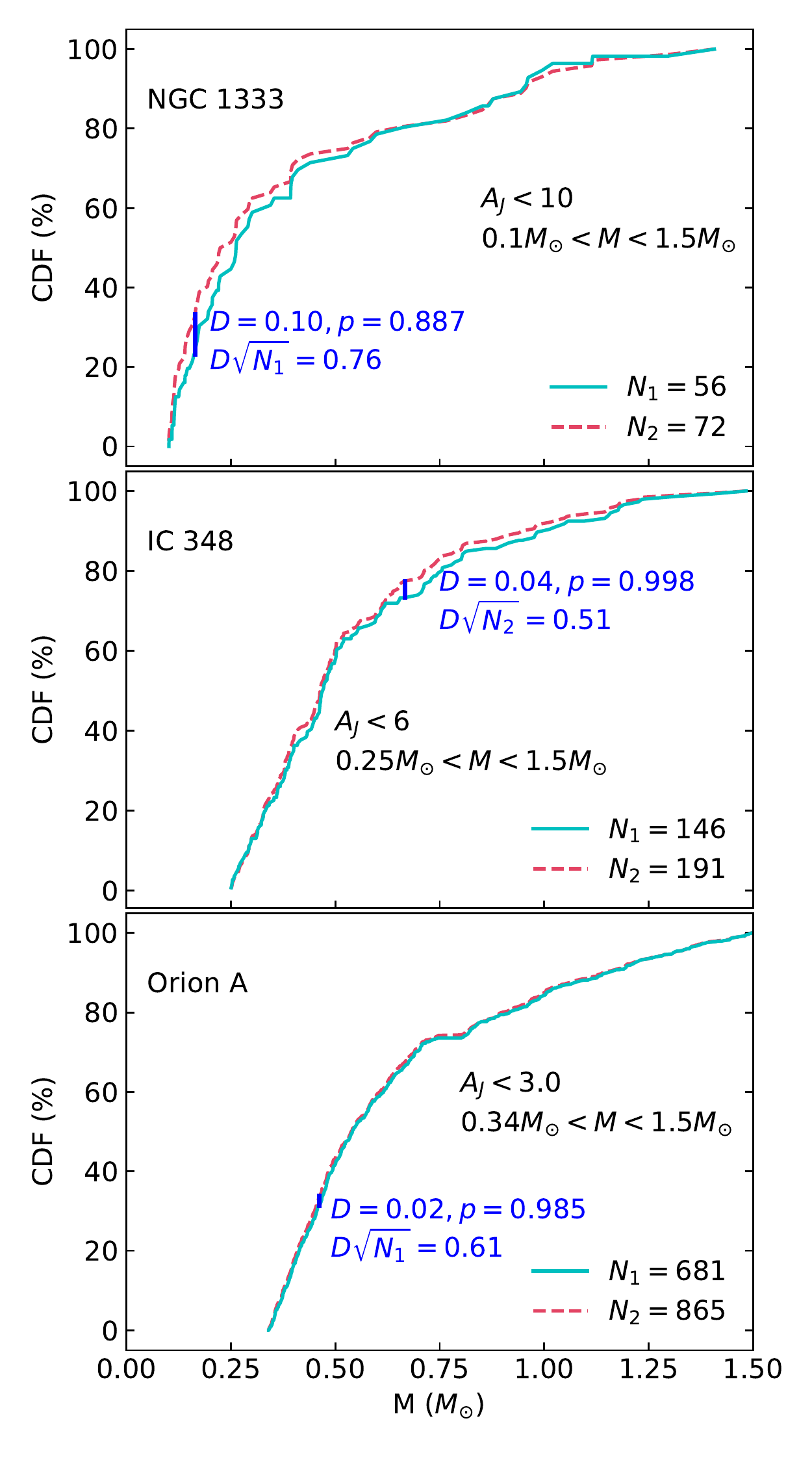}
\caption{Cumulative distribution function (CDF) of stellar masses for the samples in each region
within the representative mass ranges and $A_J$ limits.
In each panel, the dashed pink line shows CDF of the whole population, 
while the solid cyan line shows CDF of the 
used stars. The blue line segment indicates the maximum distance between two CDFs.
\label{fig:cdf}}
\end{figure}

By inspection of the left panel of Figure \ref{fig:cmd_ONC}, we can see 
that at the extinction limit of $A_J=3.0$, there are a few stars more massive than 
0.35 $M_{\odot}$ falling below the $H=12.5$ detection limit.
This may indicate that our method used to assess representativeness is, 
for some reason, yielding too low (high) limits to stellar mass (extinction).
We can check if the selection is indeed representative by 
calculating the cumulative distribution function (CDF) of stellar masses for the 
samples in each region, both the ``used'' ones (red) and the whole distributions 
(red+blue+grey+green, in Figure \ref{fig:cmd_NGC1333},
\ref{fig:cmd_IC348}, and \ref{fig:cmd_ONC}. Note that in Orion A
we have omitted all targets within 5$'$ from the center).
At the $A_J$ limit and representative mass range of each cluster, we infer the extinction
corrected $J$ band magnitude by de-reddening along the extinction vector on the 
$J$--($J-H$) panel, and convert $M_J$ to stellar mass using the BCAH98 isochrones.
The resulting CDFs are shown in Figure \ref{fig:cdf}, along with the 
number of sources in each sample ($N_1$, $N_2$).
We perform an one-dimensional K-S test for the null 
hypothesis that the two samples are drawn from the 
same continuous distribution. The returning $p$-value is quite high in all regions.

Therefore, the lower mass limits selected by the $p=0.2$ contours in Figure \ref{fig:kstest_NGC1333},
\ref{fig:kstest_IC348}, and \ref{fig:kstest_OrionA} are still considered to be acceptable.
In Section \ref{sec:mass}, we will only calculate
disk frequencies in the representative mass ranges using stars within the $A_J$ limits. 
In Section \ref{subsec:mass_OrionA} where disk fractions in Orion A are studied, we will also
investigate if the results change a lot by adopting a more restrictive extinction limit of $A_J<1$.

\subsection{Discussion of This Assessment}\label{subsec:caveat}
\subsubsection{Distance Dispersion in Each Cluster}
In the above assessment, we assume a median cluster distance for all stars in each cluster.
However, in Table \ref{tab:info}, we notice that the inter-quartile range compared to the median
distance is already large. In NGC 1333, IC 348, and Orion A,
this value is $(329.3-282.3)/300.7 = 15.6\%$, $(343.0-304.2)/324.2 = 12.0\%$, 
and $(415.1-380.0)/396.7 = 8.8\%$, respectively. In Table \ref{tab:info}, we consider effects from such a distance dispersion 
by quoting the calculated representative mass range using the 25th percentile and
75th percentile distances at the same extinction limit. The lower mass limits in IC 348 and Orion A are raised by
\mbox{0.02 $M_{\odot}$} if adopting a restrictive distance estimation (75th percentile).

\subsubsection{The Usage of 2MASS Sources} \label{subsubsec:use2mass}
We are aware of the fact that the whole population of each cluster is not red+blue+grey+green,
but how far is the total of real cluster members different from our assumption? 
Since sources marked with red, blue, and
grey have already passed the membership criteria of $Gaia$ proper motion and distance,
the fraction of contamination for them should be very low. Thence, below we only focus on
discussing the 2MASS sources shown as green pluses, for which astrometric data are not available.
Note that in Figure \ref{fig:cmd_NGC1333}, \ref{fig:cmd_IC348}, and \ref{fig:cmd_ONC}, 
the number of grey (green) pluses with inferred stellar 
mass in the corresponding representative mass ranges and extinction limits are 
4 (7), 19 (7), and 119 (32), respectively. 
Therefore, although there is a substantial amount of green pluses,
only a small fraction of them are inside of the region where 2D K-S tests are performed.

We need to take two questions into consideration:
(1) how many stars $not$ in the clusters are included in the green? (Note that 
green pluses stand for 2MASS sources not included in the Gaia catalog.)
(2) how many stars in the clusters are $not$ included in the green? 
To answer the first question, we need to consider contaminations from both background 
stars and foreground field stars. Background stars are hard to be included in the 
representative mass range of the green, because they are old, distant and reddened by the 
cloud. It is easy to imagine that on each $J$--($J-H$) plane, they generally lie on 
the left side of the cluster isochrone, or on the right side of the isochrone but below 
our adopted lower mass limit. 

Foreground stars are closer to us but suffer from little extinction, so most of them lie 
upward of (brighter than) the 1.5 $M_{\odot}$ dashed boundary, where no green plus
resides. Low mass foreground stars are easier to fall into the region where K-S tests 
are performed. However, using the \citet{1992ApJS...83..111W} Galactic star count 
model, \citet{2004AJ....127.1131W} estimated a small number of Galactic field stars 
in the densest region of NGC 1333 ($\sim$10\%). Since these authors performed a 
deeper near-infrared survey ($H\leq 16.5$) than 
ours ($H \lesssim 13$), the contamination of IN-SYNC should be less compared to 10\%. 
This is probably also true for IC 348. In Orion A, the contamination
can be potentially higher because of the relative lower extinction of this cluster. However, 
since many sources brighter than $H=12.5$ have been observed, and most field star (candidates)
are removed in Section \ref{subsubsec:membership}, we also have reasons to believe that the contamination is 
also a small fraction.

The second question is comparable to asking if any cluster members in the representative 
mass range are fainter than the 2MASS limit. The answer is probably true, in the sense 
that protostars with high extinction are harder to be detected. However, we do not think 
these sources in our sample that we classified the spectra for are actually protostars.  
That is very unlikely. However, they have such unusual infrared colors that we are not 
confident of the extinction estimates, and therefore not confident of the disk properties. 
They clearly have significant circumstellar material, probably dominated by larger 
envelope emission, but they are rare in our sample and we prefer to remove them (to some 
extent making our IR excess fractions lower limits). The removal of protostars is presented
in Section \ref{subsec:remove protostar}.

Therefore, although the assessment in Section \ref{subsec:23} is not perfect, we still consider it to be an 
acceptable way to determine the representative mass range of an extinction-limited sample.

\begin{figure}[ht!]
\includegraphics[width=\columnwidth]{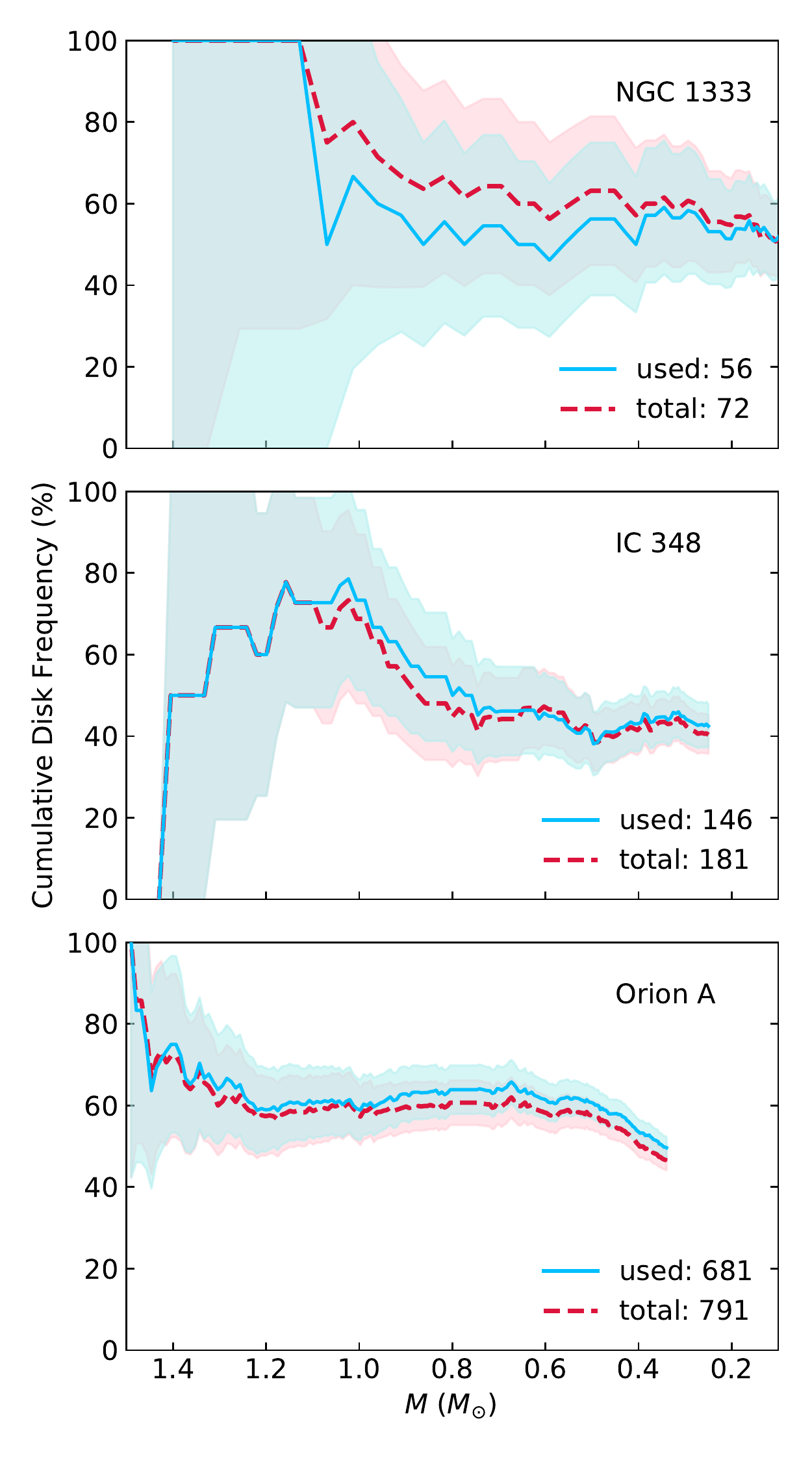}
\caption{Cumulative disk frequency for the samples in each region
within the representative mass ranges and $A_J$ limits.
In each panel, the dashed pink line shows results of the whole population, 
while the solid blue line shows results of the used stars. The light pink and light
cyan shades indicate statistical uncertainties. \label{fig:cdf2}}
\end{figure}

\subsubsection{Is there a bias towards disk-bearing YSOs?}
We have demonstrated that the finally used sample in each cluster is not biased as a function
of location on the $J$--($J-H$) CMD. Since disk properties are crucial to this work, however,
another consideration is that stars with disks are easier to be identified by previous 
membership studies. Therefore, it is possible that the used sample is biased towards sources with mid-IR excesses.
We check if this is indeed the case by considering the cumulative disk frequencies more massive
than a certain stellar mass in the representative
mass range and extinction limit. The results from the ``used'' sample and the total 
are shown as the solid blue and dashed pink line in Figure \ref{fig:cdf2}, respectively.
The background shades indicate the 1-$\sigma$ poisson 
statistical uncertainty ($\sqrt{N_{\rm disk}}/N_{\rm all}$). Please see 
Section \ref{subsec:determine excess} and \ref{subsec:classify} for the definition of ``disk frequency'' in this paper.
In principle, the number of sources in the two samples in Figure \ref{fig:cdf2} should be the same
as in Figure \ref{fig:cdf}. However, since some unused stars do not have IRAC data,
and are thus excluded from the calculation, less sources are used in creating Figure \ref{fig:cdf2}.

Statistically speaking, the cumulative disk frequency as a function of stellar mass for the
``used'' sample is consistent with the total, in the sense that the differences are all less than
1-$\sigma$. Therefore, we conclude that our samples are also representative in terms of disk properties.

\begin{figure}[ht!]
\includegraphics[width=\columnwidth]{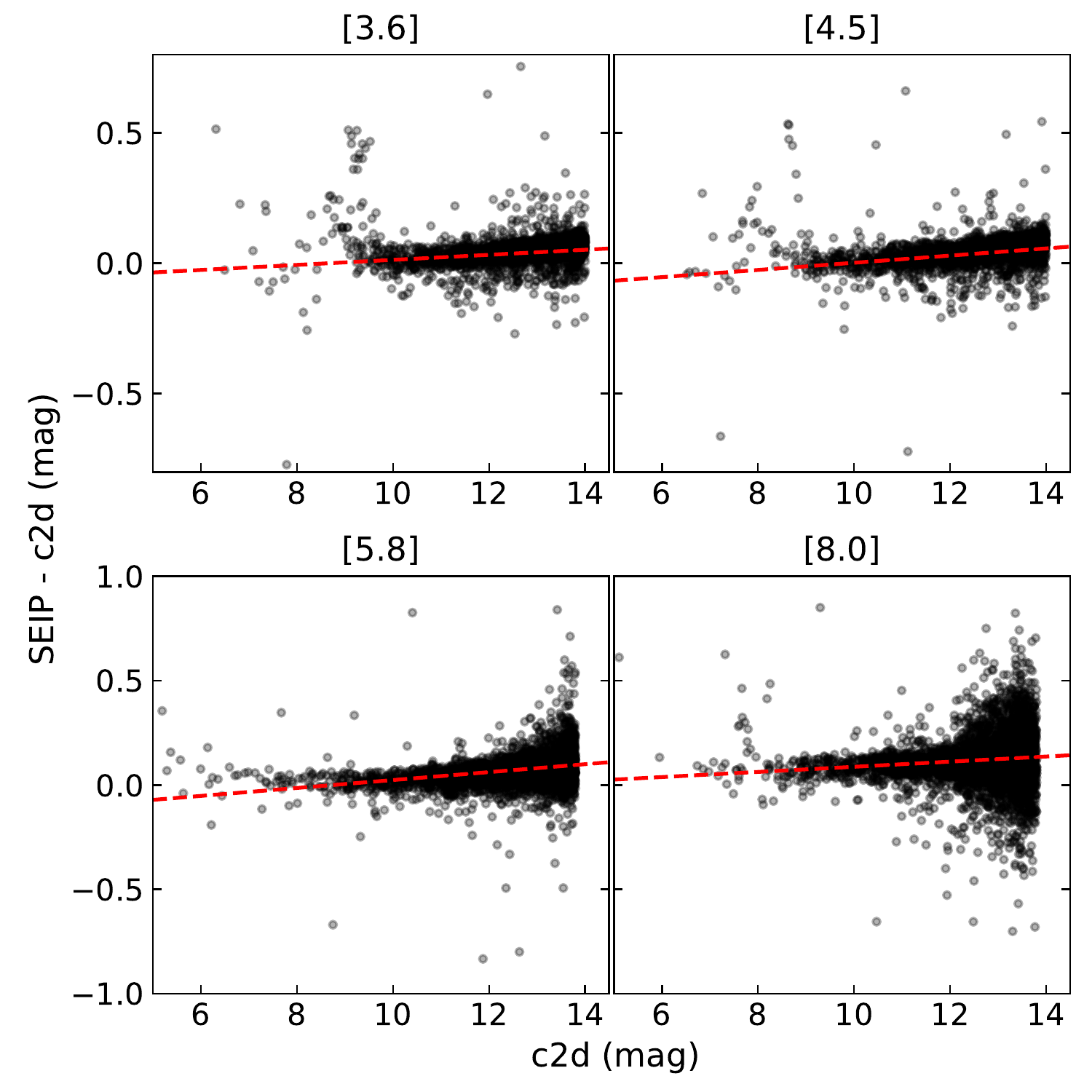}
\caption{The difference of magnitudes reported by SEIP and c2d as a function of 
the c2d magnitude for a set of sources that have been reduced by both programs.
The red lines show the fitted least squares linear regression. \label{fig:offset}}
\end{figure}

\begin{figure*}[ht!]
\includegraphics[width=\textwidth]{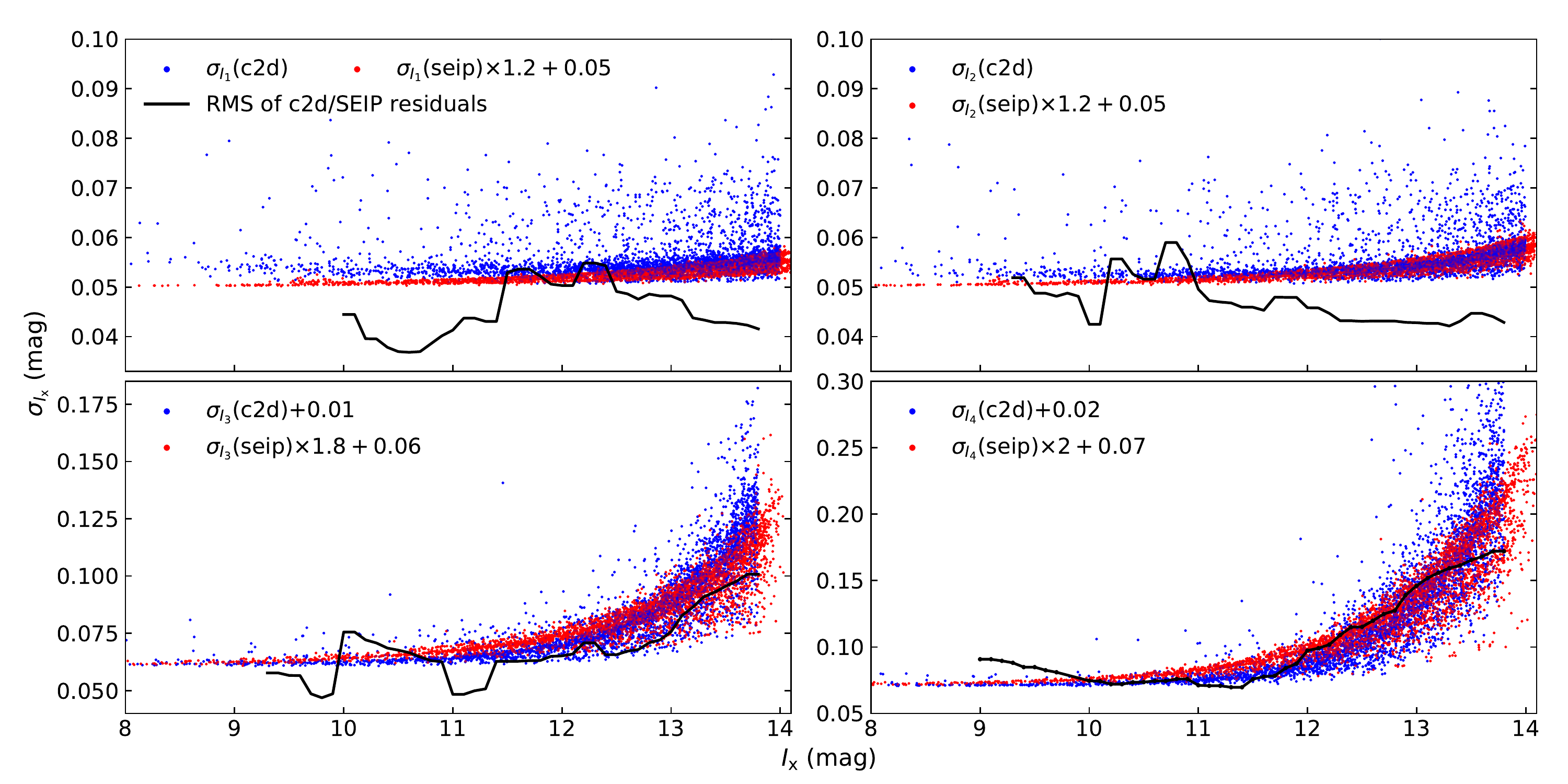}
\caption{IRAC photometric uncertainties as a function of magnitude in the four channels.
The black lines show the rms residuals of the observed scatter in Figure \ref{fig:offset}.
Blue and red dots are data from the c2d and SEIP delivery, respectively.
The legend in each panel shows how the original photometric uncertainties are increased to
be at least in the same order of the black line.
\label{fig:uncertainty}}
\end{figure*}

\section{The disk diagnostic: 4.5 $\micron$ Excess}\label{sec:diagnostic}
Widely used diagnostics to distinguish between stars with and without disks include (but 
may not be limited to): 
(1) the amount of infrared continuum excess;
(2) color-color plots constructed from combined 2MASS, WISE and $Spitzer$ photometry;
(3) infrared slope of the spectral energy distribution (SED). 
Given that the effective temperature (spectral type) are known for all stars in our 
sample, we adopt the first approach as our primary disk diagnostic. In Section 
\ref{subsec:irac}, we describe our IRAC IR data. The determination of extinctions and 
infrared excesses are outlined in Section \ref{subsec:determine excess}. Our definition of 
primordial disk is given in Section \ref{subsec:classify}. The removal of protostars is 
described in Section \ref{subsec:remove protostar}.

\subsection{IRAC Data} \label{subsec:irac}

As stated above (see Section \ref{subsubsec:photometry}), our photometric data are reduced 
by two different programs that have used different reduction techniques. To determine if 
there is a systematic difference between the flux reported by SEIP and c2d catalogs, we 
collect $\sim$30,000 sources in the Perseus region that have been reduced by both 
programs. We show the difference of their reported magnitudes as a function of the c2d 
magnitude in Figure \ref{fig:offset}. Magnitudes are converted from flux densities by 
adopting the zero magnitude flux given in the IRAC Instrument Handbook. Sources too bright 
or too faint are cut off to only consider the range of magnitudes covered by our IN-SYNC 
sample. Generally speaking, for the same sources, SEIP magnitudes are greater than that in 
c2d, especially in the two channels at 5.8 and 8.0 $\micron$. We perform an unweighted 
least squares linear regression for the IRAC four channels, as shown as red lines in Figure 
\ref{fig:offset}. To mitigate uncertainties introduced by the inhomogeneity, we add this 
offset (red lines) from the c2d magnitudes for sources in the Perseus sample. Figure 
\ref{fig:offset} also shows that there is not only a shift but also a scatter
between the photometry of c2d and SEIP. Below we explain how photometric uncertainties 
reported by c2d and SEIP are inflated to match the observed scatter in Figure 
\ref{fig:offset}. 

The IRAC photometric uncertainties reported by c2d are at least 0.05 mag, which is 
measured by the repeatability of flux measurements. Since this systematic error is not 
considered by SEIP, but is likely present in the data, we first add a floor of 0.05 mag to 
the SEIP uncertainties. After that, SEIP still underestimates flux uncertainties compared 
with c2d. Therefore, before adding the systematic error of 0.05 mag, we multiply the 
original SEIP errors by a factor of 1.2 at 3.6 and 4.5 $\micron$, a factor of 1.8 at
5.8 $\micron$ and a factor of 2.0 at 8.0  $\micron$ to bring it to the same scale with c2d.
We examine the trend of the scatter in Figure  \ref{fig:offset} as a function of  
magnitude by calculating a running rms residual with a window size of 1 mag and a step 
size of 0.1 mag. The resulting trend is shown as the black line in Figure 
\ref{fig:uncertainty}. A floor of 0.01 and 0.02 mag are further added to $\sigma_{I_3}$ 
and $\sigma_{I_4}$ to bring flux uncertainties to at least the order of the observed 
scatter in Figure \ref{fig:offset}. The finally used scale of flux uncertainties are shown 
as the blue and red dots in Figure \ref{fig:uncertainty}.

\subsection{Determination of Infrared Excess}\label{subsec:determine excess}
\begin{figure*}[ht!]
\includegraphics[width=\textwidth]{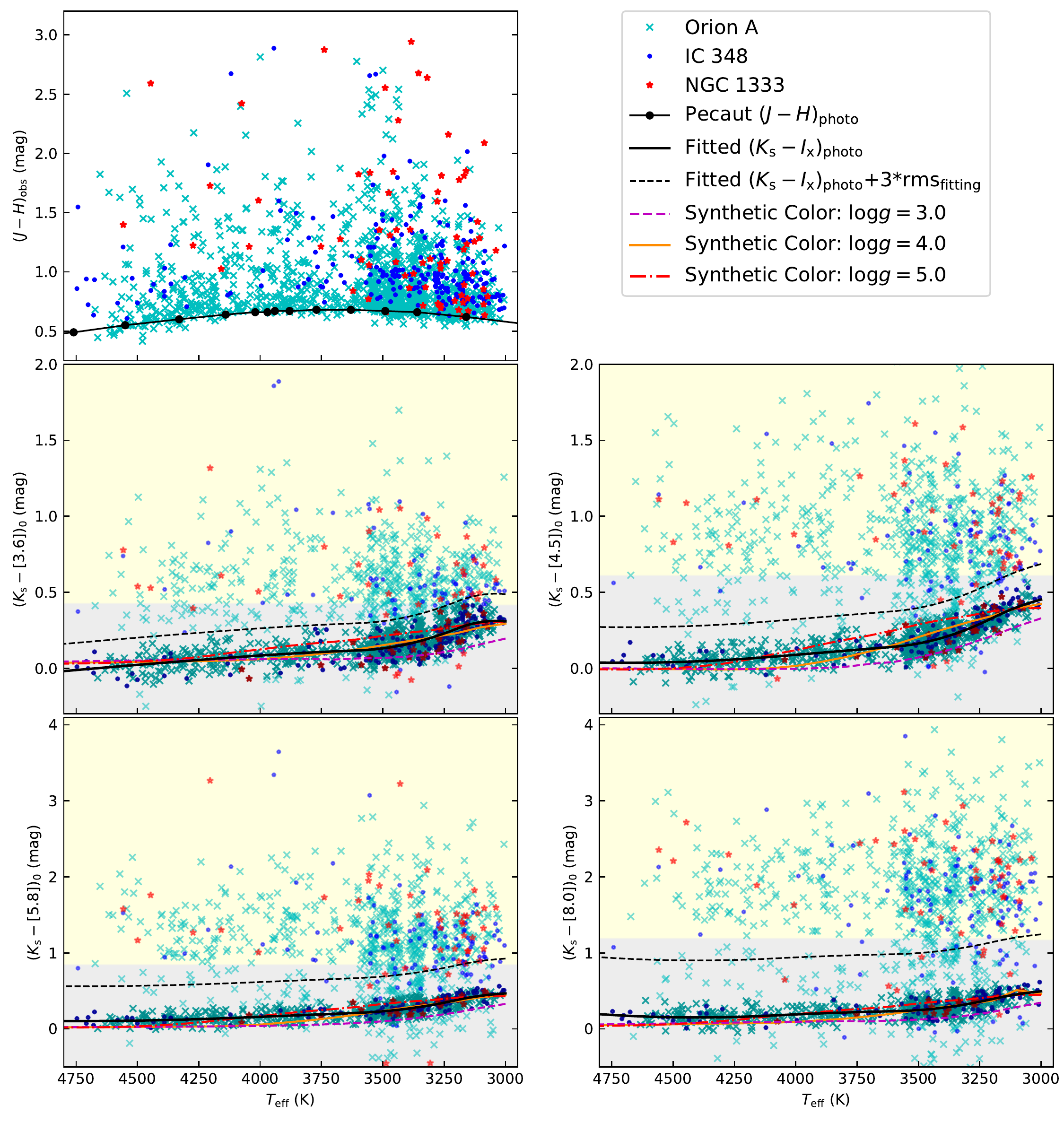}
\caption{Upper left panel: observed $(J-H)$ as a function of $T_{\rm eff}$.
The other four panels: data points are extinction corrected $(K_{s}-I_{\rm x})_{0}$ color 
as a function as $T_{\rm eff}$ for sources in our sample (those in deeper colors
are finally used to fit the color loci, see text); 
colored lines are synthetic colors calculated from spectra models;
black lines are empirical colors derived by fitting the data.
Dashed lines in the bottom four panels show 
$(K_{\rm s}-I_{\rm x})_{\rm locus}+3\sigma_{(K_{\rm s}-I_{\rm x})_{\rm locus}}$}.\label{fig:relation}
\end{figure*}

The color we observe for a YSO is a combination of effects from photospheric color, 
reddening, and intrinsic excesses. We follow the definition of intrinsic IR excess given 
by \citet{Hillenbrand1998}:
\begin{equation}\label{eq:extinction}
\Delta(J-H) = (J-H)_{0} - (J-H)_{\rm photo}
\end{equation}
where $(J-H)_{0}=(J-H)_{\rm obs} - E(J-H)$ is the extinction corrected color, and 
$(J-H)_{\rm photo}$ is the contribution of the underlying stellar photosphere.
To estimate extinction, we assume $\Delta(J-H)=0$, use the extinction law for $R_{V} = 
3.1$ from \citet{1989ApJ...345..245C} integrated over the 2MASS filters from 
\citet{2003AJ....126.1090C}, and adopt $(J-H)_{\rm photo}$ as a function of $T_{\rm eff}$
from \citet{Pecaut2013}. 

The intrinsic infrared excess is expressed by 
$\Delta(K_{\rm s}-I_{\rm x})= (K_{\rm s}-I_{\rm x})_{0} - (K_{\rm s}-I_{\rm x})_{\rm photo}$ 
(here ${\rm x}=1,2,3,4$, standing for the four IRAC channels). With the values of $A_{J}$, 
we can easily calculate $(K_{\rm s}-I_{\rm x})_{0}=(K_{\rm s}-I_{\rm x})_{\rm obs}-E(K_{\rm s}-I_{\rm x})$
by adopting the extinction law at 3--8 $\micron$ derived by \citet{2007ApJ...663.1069F}. 

To determine how $(K_{\rm s}-I_{\rm x})_{\rm photo}$ varies as a function of $T_{\rm eff}$, 
we calculate synthetic colors by convolving model spectra with the response curve of 2MASS 
and IRAC filters. Using the ``BT-Settl'' synthetic spectra \citep{Allard2012}, we compute 
colors for solar metallicity, no $\alpha$-element enhancement models with $3.0<$log$g<5.0$ 
and 3000 K $< T{\rm eff} <5000$ K. All of the stars in our sample have log$g$ and 
$T_{\rm eff}$ measurements within these ranges. In the bottom four panels of Figure 
\ref{fig:relation}, the synthetic colors are shown as the dashed magenta lines, solid 
orange lines, and dash-dotted red lines for log$g=$ 3.0, 4.0, and 5.0, respectively. 
The red asterisks, blue dots, and cyan crosses are extinction corrected colors of sources 
in our sample. We notice that there appears to be a tight locus of these data points, 
which should represent emission dominated by stellar photospheres. While the models 
describe the large scale structure of the diskless loci reasonably well, there are
discrepancies at the $\sim$0.05 mag level. To eliminate these, we derive the 
$(K_{\rm s}-I_{\rm x})_{\rm photo}$--$T_{\rm eff}$ 
relation by empirically fitting the observed loci.

To select a number of sources along the loci for fitting we follow three steps for each 
$(K_{\rm s}-I_{\rm x})_{0}$--$T_{\rm eff}$ panel: 
(1) We utilize the $K$-means clustering algorithm \citep{Lloyd1982} to partition data into 
2 groups, and retain the group with smaller $(K_{\rm s}-I_{\rm x})_{0}$ mean. In 
Figure \ref{fig:relation}, we indicate the decision boundary by assigning a background 
color to each group (light yellow and light grey). This step allows us to remove most 
outliers that are sources with excesses.
1104, 1047, 947, 844 sources are retained for the fitting of $\Delta(K_{\rm s}-I_1)$, 
$\Delta(K_{\rm s}-I_2)$, $\Delta(K_{\rm s}-I_3)$, and $\Delta(K_{\rm s}-I_4)$, respectively. 
Some sources only have IRAC data at shorter wavelength due to a loss of sensitivity
as we go to longer wavelength. 
(2) For the remaining $\sim$1000 sources, we iteratively 
fit a linear relation between $(K_{\rm s}-I_{\rm x})_{0}$
and $T_{\rm eff}$ with 3$\sigma$-clipping --- any outliers more distant than three times 
the rms of the fit residuals are removed.
(3) By eye inspection of data points on the locus,
we expect the $(K_{\rm s}-I_{\rm x})_{\rm photo}$ color slightly increase with decreasing stellar temperatures.
This should be a real feature, because synthetic colors also show this characteristic.
Therefore, we further iteratively fit a two degree spline of the remaining data with 2.5$\sigma$-clipping.
In order to prevent over-fitting, we manually set the interior knots at 3200, 3600, and 4100 K.
973, 867, 750, 626 sources are finally used for the fitting of $\Delta(K_{\rm s}-I_1)$, 
$\Delta(K_{\rm s}-I_2)$, $\Delta(K_{\rm s}-I_3)$, and $\Delta(K_{\rm s}-I_4)$, respectively. 
These sources are shown in deeper colors in the corresponding panels.

We also need to know the uncertainty in fitting the locus of diskless stars. To this end,
we consider all the sources with $(K_{\rm s}-I_{\rm x})_{\rm obs} < (K_{\rm s}-I_{\rm x})_{\rm locus}$,
i.e., those data points lying below the black lines on the bottom panels in Figure \ref{fig:relation}.
Then $\sigma_{(K_{\rm s}-I_{\rm x})_{\rm locus}}$ is evaluated by the rms of 
$(K_{\rm s}-I_{\rm x})_{\rm locus}-(K_{\rm s}-I_{\rm x})_{\rm obs}$ of these sources.
We expect that the scatter of data points $above$ the fitted lines should be very similar to that
$below$ the lines. The uncertainties are 0.060, 0.078, 0.153, and 0.250 mag
at 3.6, 4.5, 5.8, and 8.0 $\micron$, respectively. 
  
\subsection{Disk Classification}\label{subsec:classify}
Sources are thought to possess excess at a certain wavelength if
the intrinsic color exceeds the fitted relation by both 3 times the scatter of our fitting 
($\sigma_{(K_{\rm s}-I_{\rm x})_{\rm locus}}$) and 3 times the internal uncertainty associated with the 
flux ($\sigma_{(K_{\rm s}-I_{\rm x})_{0}}$), i.e.,
\begin{equation}\label{eq:excess}
\begin{aligned}
&\Delta(K_{\rm s}-I_{\rm x})>3\times \sigma_{(K_{\rm s}-I_{\rm x})_{\rm locus}}\\
&\Delta(K_{\rm s}-I_{\rm x})>3\times \sigma_{(K_{\rm s}-I_{\rm x})_{0}}
\end{aligned}
\end{equation}

\begin{figure}[ht!]
\includegraphics[width=\columnwidth]{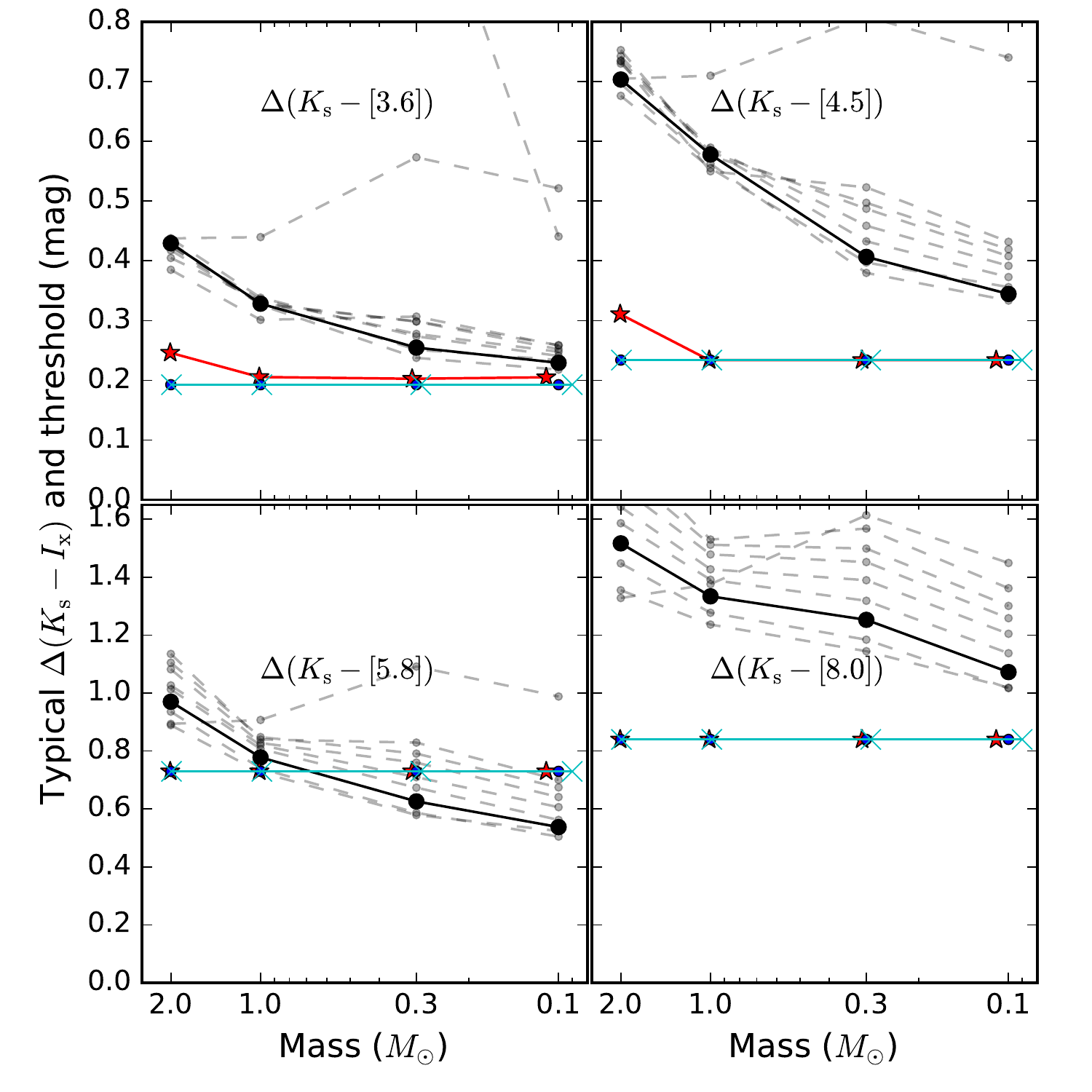}
\caption{Black: expected colors of $\Delta(K_{s}-I_x)$ if there are fiducial 
disks around young stars with stellar masses of 2.0, 1.0, 0.3, and 0.1$M_{\odot}$ (solid lines: viewed at an inclination 
of 63.26$\degree$; dashed lines: viewed at other inclinations from 18.19$\degree$ to 87.13$\degree$).
Colored lines: typical values of $max$($3\times\sigma_{(K_{\rm s}-I_{\rm x})_{\rm locus}}$, 
$3\times \sigma_{(K_{\rm s}-I_{\rm x})_{0}}$ in NGC 1333, IC 348, and Orion A are marked in
red asterisks, blue dots, and cyan crosses.
\label{fig:bias}}
\end{figure}

A crucial issue in this method is that 3.6--8.0 $\micron$ emission for a disk irradiated by a very 
faint star is coming from much smaller radii than if irradiated by a much brighter star. Therefore, there 
is an observational bias that it is harder to probe the excess for the faintest stars. 
To mitigate the effects from such a bias, we hope that even for the lowest mass stars, 
the typical amount of $\Delta(K_{\rm s}-I_{\rm x})$ should be much larger than 
both $3\times\sigma_{(K_{\rm s}-I_{\rm x})_{\rm locus}}$ and 
$3\times \sigma_{(K_{\rm s}-I_{\rm x})_{0}}$. Is this really the case?

To answer this question we make use of disk SEDs computed by \citet{2006ApJS..167..256R}.
In Figure \ref{fig:bias}, we show in black lines the amount of excess that we expect to see 
if there are fiducial primordial disks around stars of four different masses.
Fiducial disks are selected from a grid of SEDs by requiring age = 1--3 Myr, 
inner disk radius = sublimation radius, etc. In colored lines we show typical values of 
$max$($3\times\sigma_{(K_{\rm s}-I_{\rm x})_{\rm locus}}$, 
$3\times \sigma_{(K_{\rm s}-I_{\rm x})_{0}}$) in the three clusters.

As demonstrated in Figure \ref{fig:bias},
the typical values of excess threshold, $max$($3\times\sigma_{(K_{\rm s}-I_{\rm x})_{\rm locus}}$, 
$3\times \sigma_{(K_{\rm s}-I_{\rm x})_{0}}$), is as large 
as (and sometimes larger than) the amount of excess at 3.6 and 5.8 $\micron$.
However, at 4.5 and 8.0 $\micron$, a fiducial disk should be selected by our criterion,
since the amount of excess exceeds that of the threshold.
In this paper, we define stars with primordial disks as those possess excess at 4.5 $\micron$.

\subsection{Removing Protostars}\label{subsec:remove protostar}
The detection of infrared excesses can be attributed to the existence of either envelopes or disks. 
Protostars (flat spectrum, Class 0 or Class 1 sources) with dusty envelopes should be removed 
before the calculation of disk frequencies. Most experimental criteria used to identify protostars made use
of IRAC 3.6, 4.5, and Multiband Imaging 
Photometer for $Spitzer$ (MIPS) photometry at 24 $\micron$.
See \citet[their Eq.1 and Eq.2]{2012AJ....144...31K} for an example. 
However, since a few Perseus objects and most Orion sources don't have MIPS data in 
c2d or SEIP (because of bright nebulosity, saturation, and SEIP's strict cuts of extended sources), we 
use the following equation to remove protostars:
\begin{equation}
(I_1 - I_2)_0 > 0.7
\end{equation}
Envelope models predicte that the above criterion \citep{2004ApJS..154..363A} can separate most
protostars from disk-bearing stars.

4 (5.4\% in NGC 1333), 4 (1.6\% in IC 348), and 11 (0.8\% in Orion A) sources are removed from the three clusters.
After that, we are left with 70 sources in NGC 1333, 239 in IC 348, and 1332 in
Orion A. A table of these sources is provided in Appendix.

\section{Disk Frequency and Stellar Mass}\label{sec:mass}
In this section we estimate the frequency of primordial disks for each cluster, and 
discuss evidence of trends in disk frequency over the representive mass range (Table 
\ref{tab:info}). In Section \ref{subsec:intermediate_mass}, we compile
a list of intermediate mass stars in each region. The dependence of 
disk frequency on stellar mass is studied in Section \ref{subsec:mass_NGC1333},
\ref{subsec:mass_IC348}, and \ref{subsec:mass_OrionA} for NGC 1333, IC 348, and Orion
A, respectively. The results in the first two clusters are compared with the recent work of
\citet{Luhman2016} in Section \ref{subsec:compare_Luhman2016}.

\subsection{Compiling a List of Intermediate-mass Stars} \label{subsec:intermediate_mass}
To further compare disk frequencies of our low-mass sample with that of higher mass stars, 
we search from literature for intermediate-mass stars (2.2--5 $M_{\odot}$). It is probable 
that focused membership studies can provide us with a complete sample of these early-type 
stars due to their brightness. The spectral types corresponding to this mass range in 
each cluster are outlined in Table \ref{tab:info} based on isochrone models by 
\citet{2000A&A...358..593S} (BCAH98 did not compute isochrones for $M>1.5$ $M_{\odot}$ 
stars).

We obtain 4 and 28 intermediate-mass stars in NGC 1333 and IC 348 from 
\citet{Luhman2016}, all of which have c2d $Spitzer$ IRAC photometry except for 
the G0 type star 2MASS 03443200+3211439 in IC 348. We also remove the A3 type star SSTc2d 
J034432.0+321144 in IC 348 for its lack of 2MASS photometry. For Orion A, we obtain 55 
intermediate-mass stars in ONC from \citet{2012ApJ...748...14D}, and 74 in L1641 from 
\citet{2012ApJ...752...59H, 2013ApJ...764..114H}. 89 (29+60) of them have extracted SEIP 
IRAC fluxes at 4.5 $\micron$.

Extinction for these intermediate-mass stars are estimated in the same way as outlined in 
Section \ref{subsec:determine excess}, but using the temperature (spectral type) scale for
main sequence stars from \citet{Pecaut2013}. Their photospheric colors
$(K_{\rm s}-I_{\rm x})_{\rm photo}$ are assumed to be the same as a 1.5 $M_{\odot}$ star. It 
is evident from Figure \ref{fig:relation} and synthetic photometry
that the loci do not vary greatly for stars more massive than 1.5 $M_{\odot}$. 
Two of the 89 stars in Orion A exhibit $(I_1-I_2)_{0}>0.7$ and are removed as protostars.

Now there are 4, 26, and 87 intermediate-mass stars in NGC 1333, IC 348, and Orion A, respectively.
Similar to what we have done in the compilation of our low-mass star sample, we also
apply the same proper motion and distance constraints outlined in 
Section \ref{subsubsec:membership} on these intermediate-mass stars, if $Gaia$ data is available.
After that, we are left with 4, 24, and 65 intermediate-mass stars in NGC 1333, IC 348, and Orion A.

\subsection{NGC 1333} \label{subsec:mass_NGC1333}
\begin{figure}[ht!]
\includegraphics[width=\columnwidth]{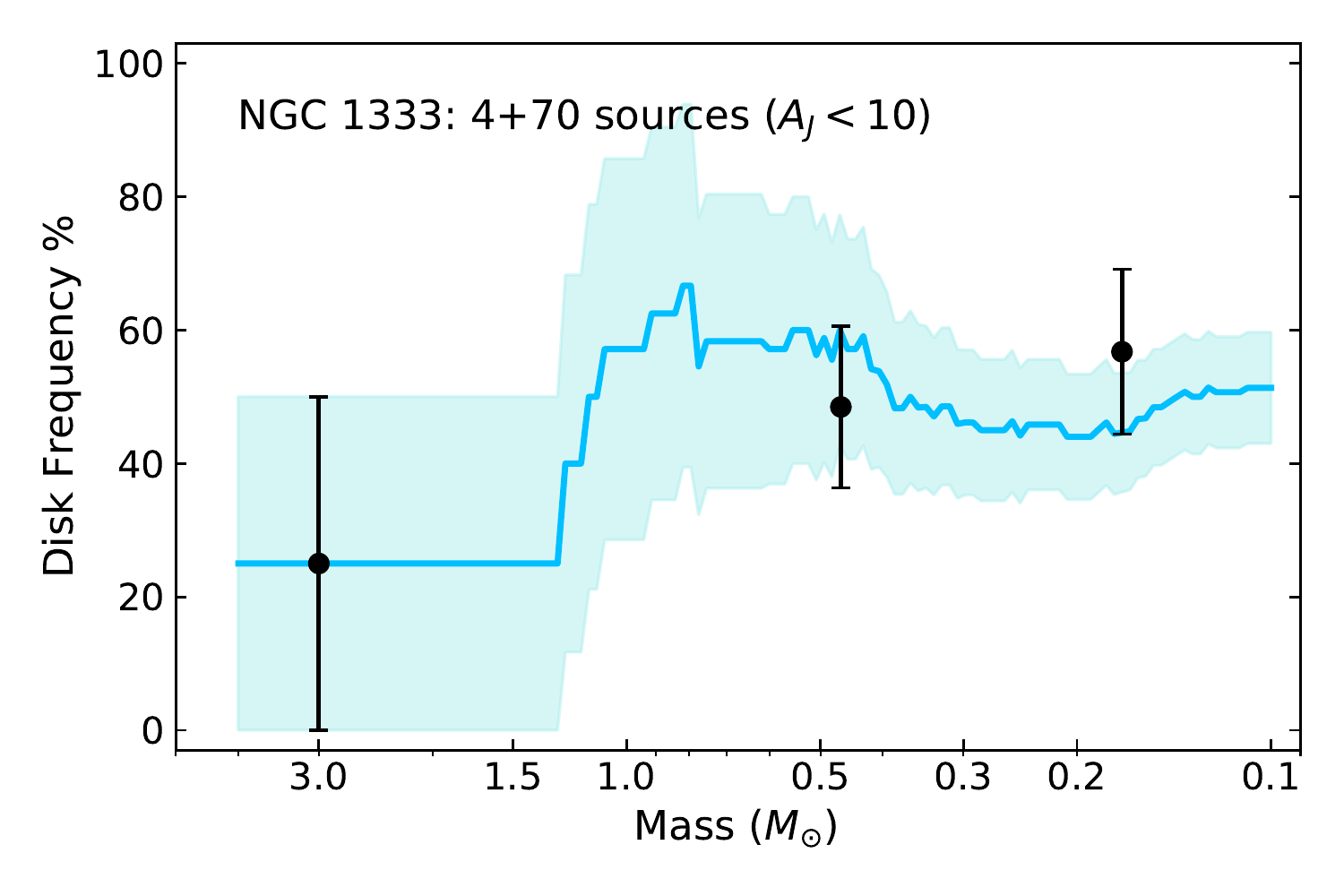} 
\caption{Black: disk frequency as a function of stellar mass in NGC 1333. 
Blue line: cumulative disk frequency derived by considering all 
stars more massive than a given mass ($X$-axis).\label{fig:freq_NGC1333}}
\end{figure}

Assuming an average age of 1 Myr for NGC 1333, we convert $T_{\rm eff}$ into mass using 
the BCAH98 isochrone. After that, we separate our sample into two mass 
bins so that each bin contains 35 sources. Figure \ref{fig:freq_NGC1333} 
demonstrates the derived disk frequency versus stellar mass. The blue line shows 
cumulative disk frequency as defined by the fraction of disks for all stars with stellar 
mass higher than a certain value. We have too few stars in the intermediate-mass range to 
allow robust comparison with low-mass stars. In the low-mass regime, our result is 
consistent with no mass dependence of disk fraction. 

The derived overall disk frequency (0.1--1.5 $M_{\odot}$, 52.9$\pm$8.7\%) is much lower than 
the 82.8$\pm$9.8\% (72/87) given in \citet[$M>0.08$ $M_{\odot}$, 
$A_{K_{\rm s}}<2$]{2008ApJ...674..336G}. The 
latter disk frequency is examined by taking the ratio of $Spitzer$-identified YSOs with 
$K_{\rm s}<14$ in the ``Main'' cluster region over all 2MASS sources down to $K_{\rm 
s}=14$ after correcting for field-star contamination, which means that disk frequency is 
calculated as N(0/I+II)/N(0/I+II+III). Therefore, the higher disk frequency in NGC 1333 
reported by \citet{2008ApJ...674..336G} can be caused by the facts that: (1) they count 
protostars, which are removed from our analysis; (2) their magnitude limit of $K_{\rm 
s}<14$ can be biased against diskless stars, since disk emission can already be prominent 
at 2 $\micron$ under circumstances of low inclination angle, high accretion rate, flared 
disk geometry, or small inner disk hole size \citep{Hillenbrand1998}.


\subsection{IC 348} \label{subsec:mass_IC348}
\begin{figure}[ht!]
\includegraphics[width=\columnwidth]{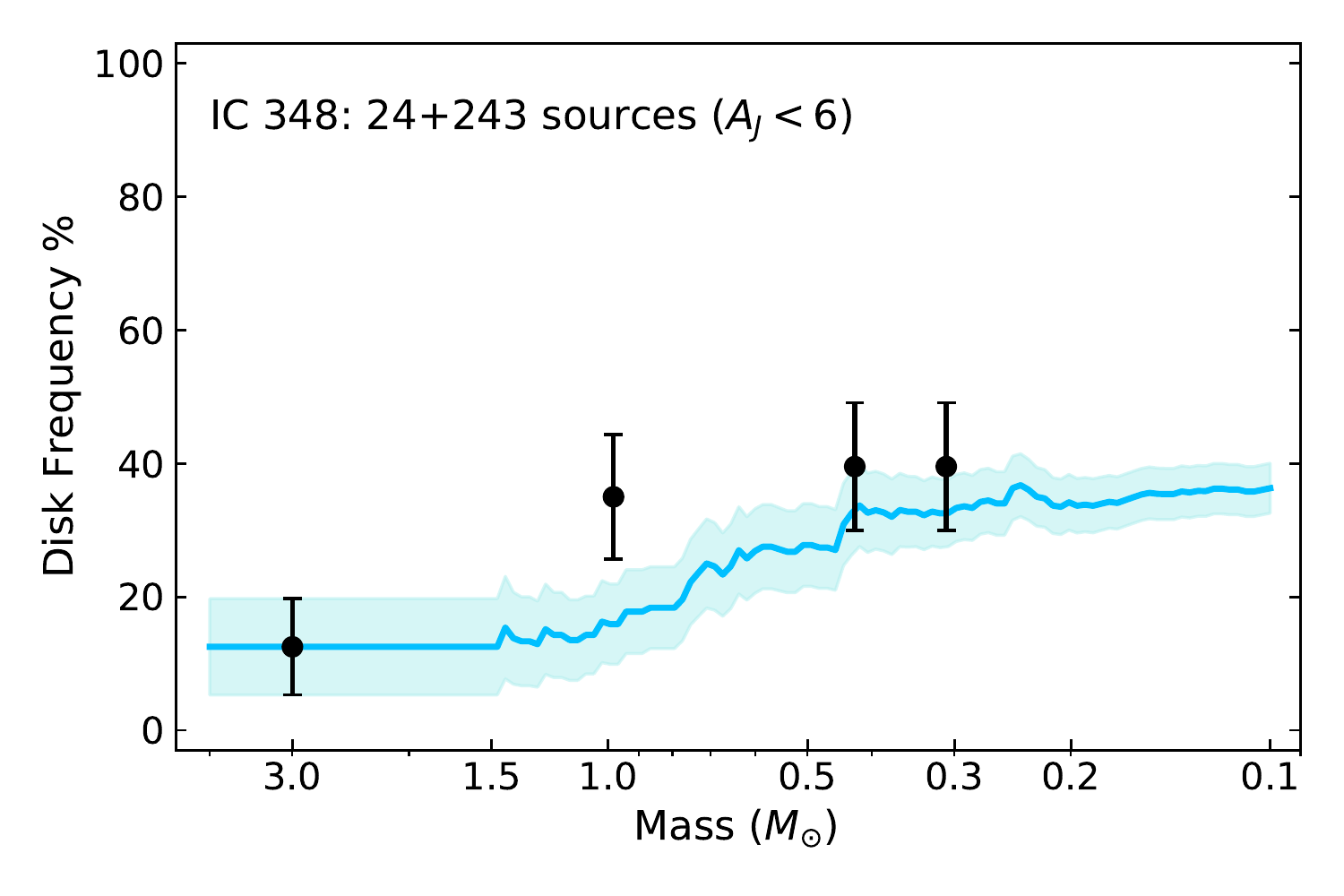}
\caption{Disk frequency as a function of stellar mass in IC 348. 
We only bin stars more massive than $0.25M_{\odot}$.
 See Figure \ref{fig:freq_NGC1333}. \label{fig:freq_IC348}}
\end{figure}

\begin{figure}[ht!]
\includegraphics[width=\columnwidth]{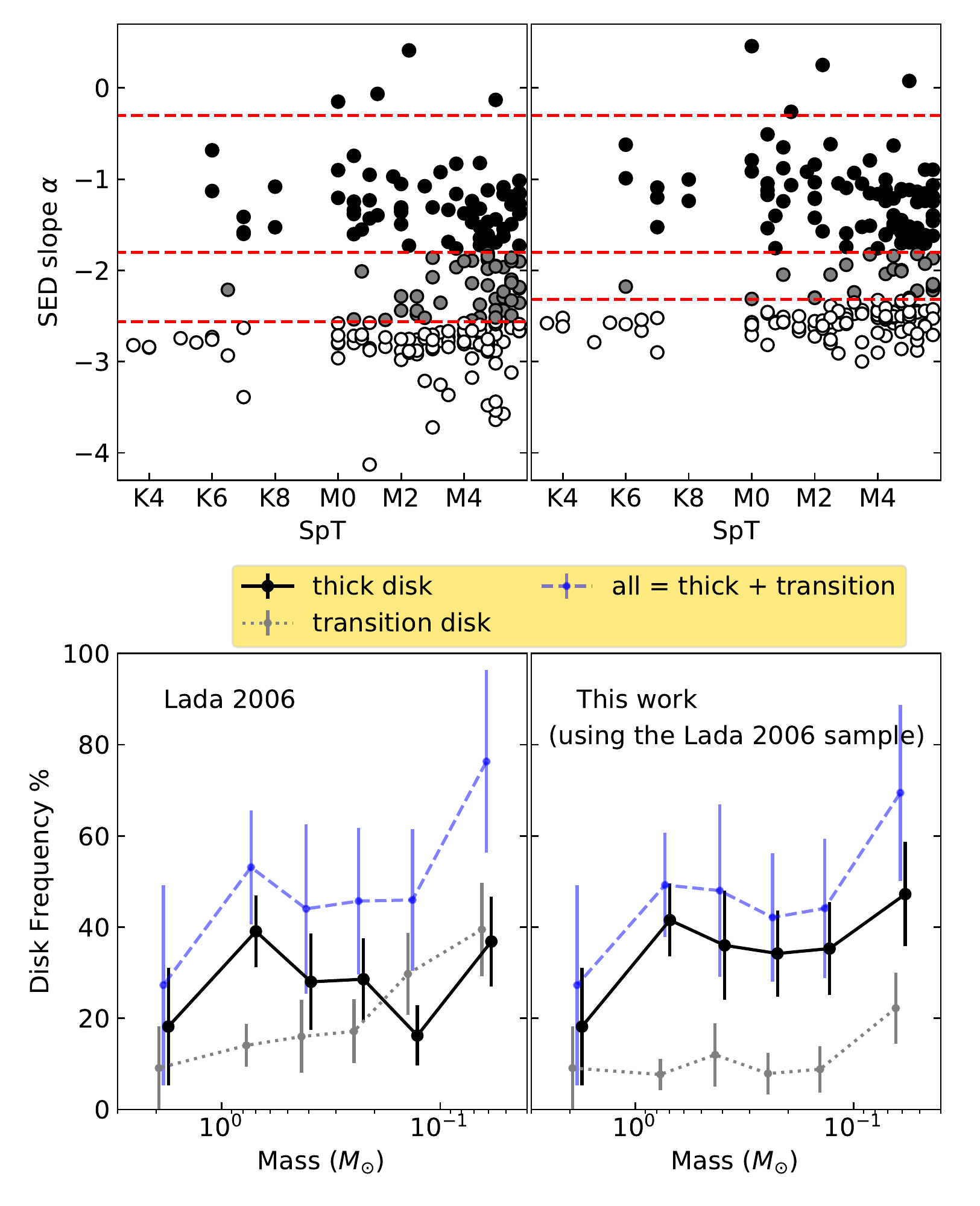} 
\caption{Upper Left: 3.6--8.0 $\micron$ de-reddened SED slope ($\alpha$) determined by Lada06. 
$\alpha$ is defined in the $\lambda f_{\lambda}$ unit (Eq. \ref{eq:alpha}).
Dashed red lines ($\alpha=-2.56, -1.8, -0.3$) show the separation of 
different disk properties adopted by Lada06.
Upper Right: 3.6--8.0$\micron$ de-reddened SED slope $\alpha$ as a function of spectral type. 
$\alpha$ are determined using c2d data and new estimation 
of extinction (Section \ref{subsec:determine excess}). Thick, transition, and diskless 
stars are color-coded in black, grey, and white, respectively. Dashed red lines are 
$\alpha=-2.32, -1.8, -0.3$. Bottom Left \& Bottom Right: Disk frequency as a function of stellar mass. 
\label{fig:cccc}}
\end{figure}

Figure \ref{fig:freq_IC348} shows disk frequency versus stellar mass in IC 348. The lowest 
disk frequency is in the highest mass bin (2.2--5.0 $M_{\odot}$). As we go to lower mass 
stars there is an increase of disk frequency. For low-mass stars (0.25--1.5 $M_{\odot}$), 
disk frequency is consistent with little dependence on stellar mass.

Lada06 also analyzed a sample of $\sim$300 known members in IC 348, and found four 
contiguous drops of disk fraction as one went to lower mass stars from 1.0 $M_{\odot}$ to 
0.1 $M_{\odot}$. It is worth checking why such a definite decline is not seen in our data. 
To this end, we start with the 234 stars with spectral type between K3 and M6 in Lada06.
For 214 stars that have data from $Spitzer$ GTO program at all IRAC bands, we show their 
de-reddened SED slope measured by Lada06 as a function of spectral type in the upper left 
panel of Figure \ref{fig:cccc}, and the resulting disk frequency in the bottom 
left panel of Figure \ref{fig:cccc}. The spectral index (SED slope) is defined by
\begin{equation}\label{eq:alpha}
\alpha = \frac{{\rm d \ lg(}\lambda f_{\lambda})}{{\rm d \ lg (}\lambda)}
\end{equation}
and is measured by a simple power-law, least squares fit to the four IRAC bands.

\begin{figure}[ht!]
\includegraphics[width=\columnwidth]{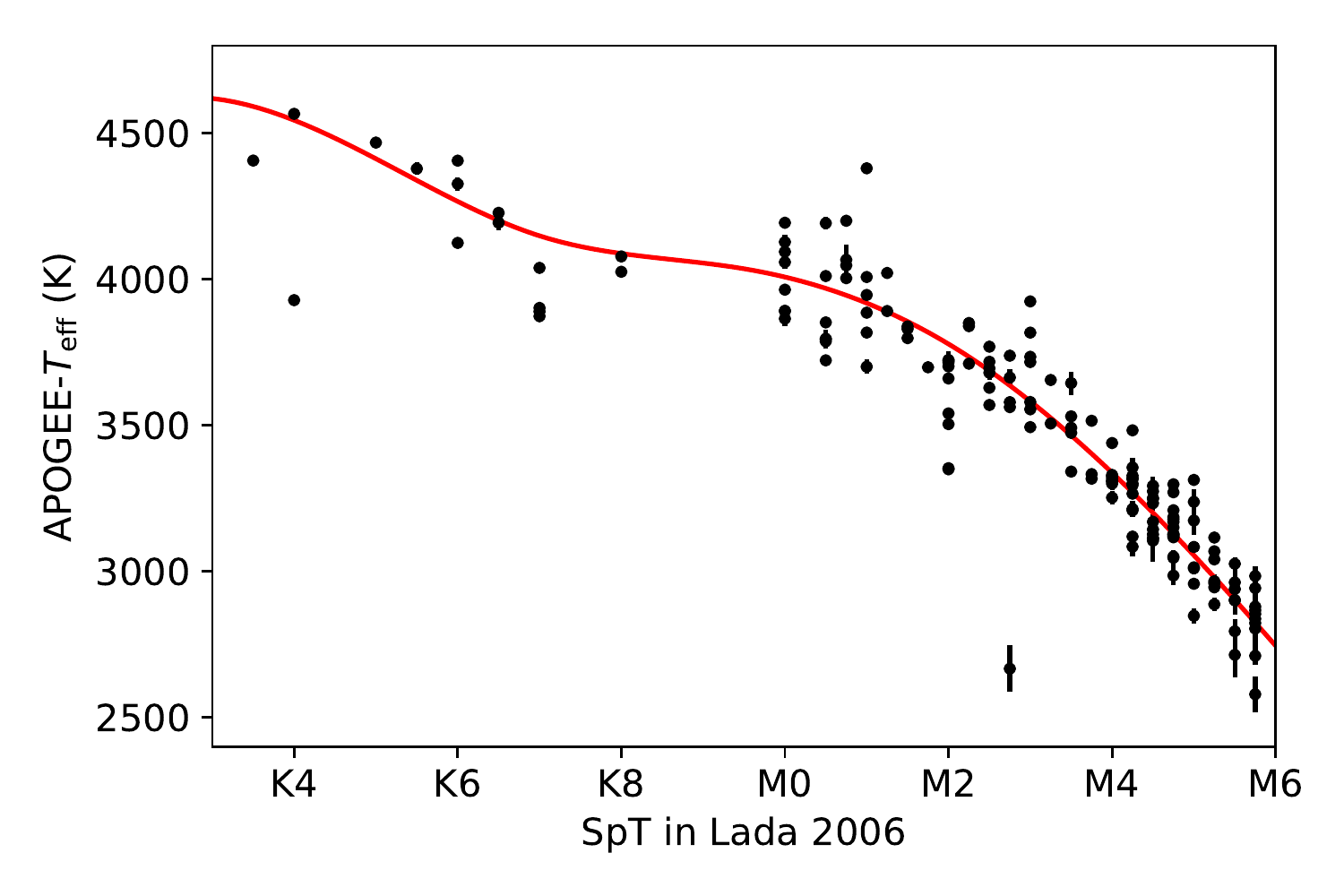}   
\caption{$T_{\rm eff}$ measured from APOGEE spectra as a function of spectral type used in 
Lada06. Only 166 stars with $\sigma_{T_{\rm eff}}<80$ K are used in the fitting.
\label{fig:convert_IC348}}
\end{figure}

In the above analysis, spectral types, which are originally from 
\citet{2003ApJ...593.1093L}, are converted into effective temperatures by fitting a 
APOGEE-$T_{\rm eff}$--SpT spline function using a set of 166 stars that have also been 
observed by IN-SYNC. The fitted relation is shown as the red line in Figure 
\ref{fig:convert_IC348}. $T_{\rm eff}$ is then converted to stellar mass using 
the BCAH98 3 Myr model. Four protostars are removed by cutting sources with $\alpha>-0.3$ 
from analysis \citep{1994ApJ...434..614G, 2015AJ....150...40Y}. By and large this is just 
a repetition of the work of Lada06 with minor adjustments.

Apart from the different adopted disk diagnostics (this work: 4.5 $\micron$ excess; Lada06: 
spectral index $\alpha$), their IRAC data is from $Spitzer$ GTO program, which is not exactly the 
same as being published by c2d (different reduction procedures can be a reason).
We noticed that there are several diskless stars having extremely steep slopes ($\alpha<-3$),
whose magnitude at 8.0 $\micron$ used by Lada06 can be more than $\sim$1 mag fainter than 
that reported by c2d. The photometric uncertainties reported by c2d are generally smaller at 8.0 $\micron$.
Furthermore, Lada06 used different methodology to estimate extinction, which
may produce some systematic differences. In general, most of stars have $A_{J}$ 
in Lada06 lower than our estimates. 

These issues motivate us to re-determine extinction for the 234 stars with K$3<$ SpT $<$ M6 
by the method outlined in Section \ref{subsec:determine excess}, and re-fit the 
de-reddened 3.6--8.0 $\micron$ SED slope for 213 stars that have c2d data at all IRAC 
bands. The newly-determined $\alpha$ is shown as a function of spectral type in the upper 
right panel of Figure \ref{fig:cccc}. 

Compared with the upper left panel of Figure \ref{fig:cccc}, where there is 
not a very distinct separation at $\alpha=-2.56$ between diskless stars and transition 
disks\footnote{Transition disk is termed as ``anemic'' disk in Lada06.}, the distribution 
of $\alpha$ for Class III stars determined by this work becomes tighter.
We follow Lada06 by classifying all sources with $\alpha > -1.8$ as primordial
disks, but change the maximum $\alpha$ for Class III objects to be $-2.32$.
The resulting disk frequencies are shown in the lower right panel of Figure \ref{fig:cccc}.
The general trend of our new disk frequencies are not sensitive to the new cutoff of $\alpha=-2.32$.
A lot of sources termed as stars with ``transition disks'' are classified as stars with optically-thick disks in our re-analysis. 
\citet{2007ApJ...662.1067H} also noted that the number of transition disks in Lada06
could be overestimated for the lowest mass stars.

In conclusion, both Figure \ref{fig:freq_IC348} and the lower right panel of Figure \ref{fig:cccc} demonstrate
that in IC 348, primordial disk frequency is lower around sources more massive than 1 $M_{\odot}$.
But we do not observe a significant drop of disk frequency at 0.25--0.4 $M_{\odot}$.
This is also consistent with the result of \citet{Luhman2008}.


\subsection{Orion A} \label{subsec:mass_OrionA}
\begin{figure}[ht!]
\includegraphics[width=\columnwidth]{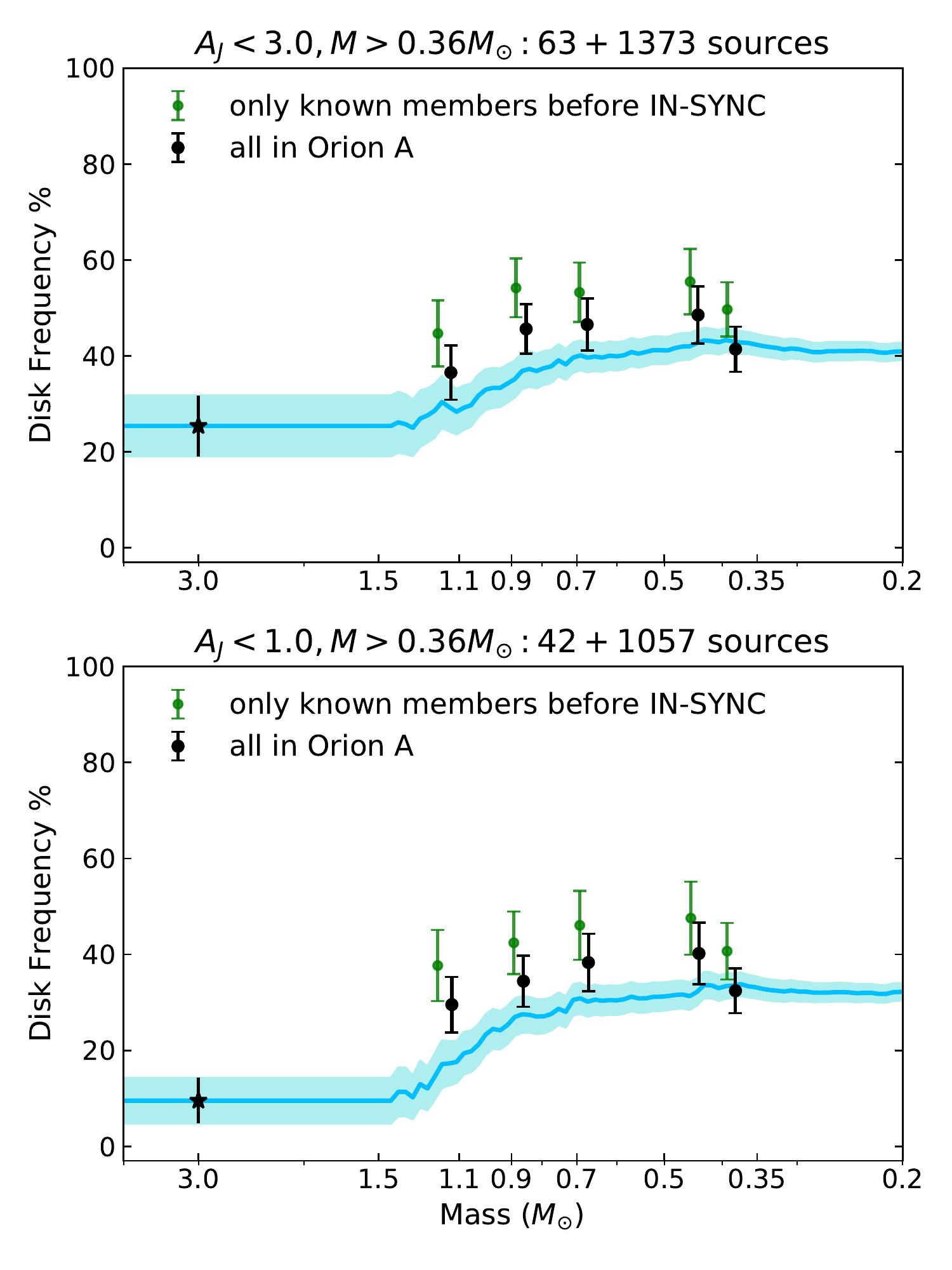}
\caption{Disk frequencies as a function of stellar mass in the Orion A molecular cluster. 
We only bin stars more massive than 0.36$M_{\odot}$.
The green data points present the results if only considering known members 
before the IN-SYNC study, which confirmed many non-disk-bearing stars. The upper panel
considers 1373 low mass sources and 63 intermediate mass sources
with $A_J<3.0$, while the lower panel only takes the 1057+42 sources
with $A_J<1.0$ into consideration. See Figure \ref{fig:freq_NGC1333}. \label{fig:freq_OrionA}}
\end{figure}

The upper panel of Figure \ref{fig:freq_OrionA} shows primordial disk frequency versus stellar mass in Orion 
A. $T_{\rm eff}$ is converted to stellar mass using the BCAH98 2 Myr model. 
Results for the lowest mass stars are not shown since our sample is not representative for 
$M<0.36$ $M_{\odot}$ (Table \ref{tab:info}, we adopt a restrictive estimate of the lower mass limit). 
The upper panel considers the $63+1373$ (intermediate-mass + low-mass) sources with $A_J<3.0$.
From $M\approx 5$ $M_{\odot}$ to $M\approx0.9$ $M_{\odot}$, there is an evident increase of disk frequency, 
which is consistent with the notion that disks live longer around lower mass stars. In the 
mass bin of 0.36--0.42 $M_{\odot}$, we observe a 6.5\% drop of disk 
frequency compare with the bin of 0.42--0.5 $M_{\odot}$. 
This decline is even more significant (beyond 1-$\sigma$ uncertainty) if the lowest mass bin is
chosen at 0.36--0.4 $M_{\odot}$.

Disk frequencies in sub-regions in the Orion A molecular cloud have also been investigated by 
previous studies. \citet{2013ApJS..207....5F} diagnosed disk properties for $\sim$1000 
sources in L1641 using 2MASS-$Spitzer$ SED slopes. Among three mass bins of 
0.1--0.32 $M_{\odot}$, 0.32--1 $M_{\odot}$, and $>1$ $M_{\odot}$, the highest disk frequency 
was found at 0.32--1 $M_{\odot}$.
{Below we consider reasons that may produce the decline of 
the observed primordial disk frequency for the lowest mass stars. }

A straightforward explanation, as mentioned in Section \ref{subsubsec:data OrionA}, is that our 
method to assess representativeness underestimates (overestimates) the lower mass limit (extinction limit)
for which our sample can be considered representative of the cluster population. 
Inspection of the left panel of Figure \ref{fig:cmd_ONC} reveals that 
at the extinction limit of $A_J=1.0$, no cluster members more massive than 0.35 $M_{\odot}$ would
drop below the detection limit of $H=12.5$. Therefore, we further study disk frequencies of 
stars with $A_J<1.0$, which gives us the lower panel of Figure \ref{fig:freq_OrionA}.
Disk frequencies in the lower panel are generally lower than that in the upper panel, 
because stars with primordial disks already have an excess at $H$ band,
which can render $\Delta(J-H)$ larger than 0. 
Therefore, our estimation of $A_J$ for disk-bearing stars
is larger than the true value. By requiring $A_J<1.0$ we are excluding more disk-bearing 
stars than diskless stars. 
In the lower panel of Figure \ref{fig:freq_OrionA},
the decline of primordial disk fraction at the lowest mass bin is still present.
We conclude that the disk fractions we measure have not been biased to lower 
values by adopting an overly permissive extinction limit for our representative sample.

\begin{figure}[ht!]
\includegraphics[width=\columnwidth]{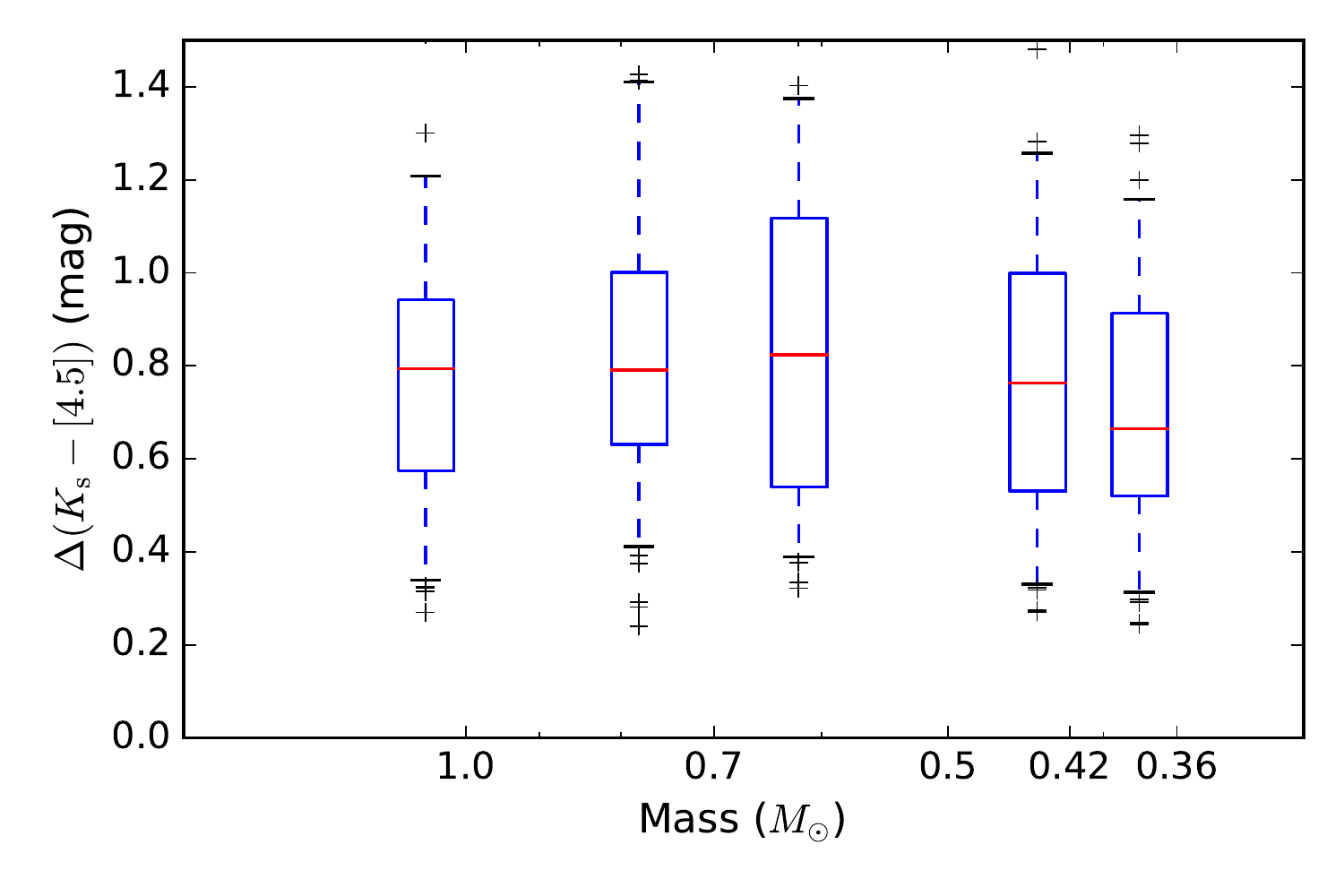}
\caption{Box-and-whisker plot of $\Delta(K_s-[4.5])$ for sources with disks in Orion A
at the extinction limit of $A_J=3.0$. The box extends from the lower to upper 
quartile values, with a red line at the median. 
The whiskers extend from the box to show the range from 5th to 95th percentile of the data. 
Outliers are marked as plus symbols. \label{fig:box_whisker}}
\end{figure}

Another explanation is large degrees of dust settling (flattened geometry) or grain growth 
\citep{2006ApJ...638..314D, 2007ApJ...662.1067H}. Both effects will introduce
an observational bias against the detection of gas-rich disk for low mass stars. In Figure 
\ref{fig:box_whisker}, we present the distribution of 4.5 $\micron$ excess for sources 
identified as having disks by Eq. (\ref{eq:excess}). Indeed the amount of excess decreases 
for stars in the mass range of \mbox{0.36--0.42 $M_{\odot}$}. Under scenarios of even the 
same degrees of dust settling or grain growth for stars of different masses, disks around 
low mass stars are easier to drop below our detection limit given that the amount of 
fiducial excess for a 0.3 $M_{\odot}$ star is already smaller than a 1 $M_{\odot}$ star (see 
Figure \ref{fig:bias}). Therefore, we are unable to claim whether dust settling operates
faster for low-mass stars, the question of which involves strong assumption on disk 
structure and physics. However, if this is indeed the case, as suggested by Lada06 and 
\citet{2007ApJ...662.1067H}, it could also be affecting NGC 1333 and IC 348, even if no 
drop is apparent in the disk fractions.

\subsection{Comparison with \citet{Luhman2016}} \label{subsec:compare_Luhman2016}
The recent census of NGC 1333 and IC 348 \citep{Luhman2016} 
also investigates disk fractions as a function of spectral type.
The authors collect $Spitzer$ IR photometry from a various of literature and 
measure disk fractions as $N$(II)/$N$(II+III). It is not explicitly specified 
whether the presence or absence of mid-IR excess is measured based on SED slope
or excess colors. Their results (Figure 23 of \citet{Luhman2016}) also show a higher disk fraction
in NGC 1333, but their disk frequencies in both clusters are slightly higher
than ours. The reason may originate from the fact that we have different definitions of ``disk frequency''.
In this paper, we are only detecting optically-thick inner disks with 4.5 $\micron$ excess.
For a few stars with only excess at $I_3$ or $I_4$, but not at $I_1$
or $I_2$, they are also counted as stars with disks in \citet{Luhman2016}. 

\section{Disk Frequency and Stellar Age}\label{sec:age}
In this section, we study a dependence of disk evolution --- as probed by disk frequencies 
--- as a function of stellar age. As an age indicator, surface gravity 
should be independent of infrared excess (at least to first order), and is thus 
independent of our classification of disk type. For this analysis, we further exclude any stars 
in our sample with $A_J$ higher than the extinction limit,
$\sigma_{{\rm log}g}>0.2$ dex, or $\sigma_{T_{\rm eff}}>120$ K,
leaving 57 sources in NGC 1333, 221 in IC 348, and 1200 in Orion A.

\begin{figure}[ht!]
\includegraphics[width=\columnwidth]{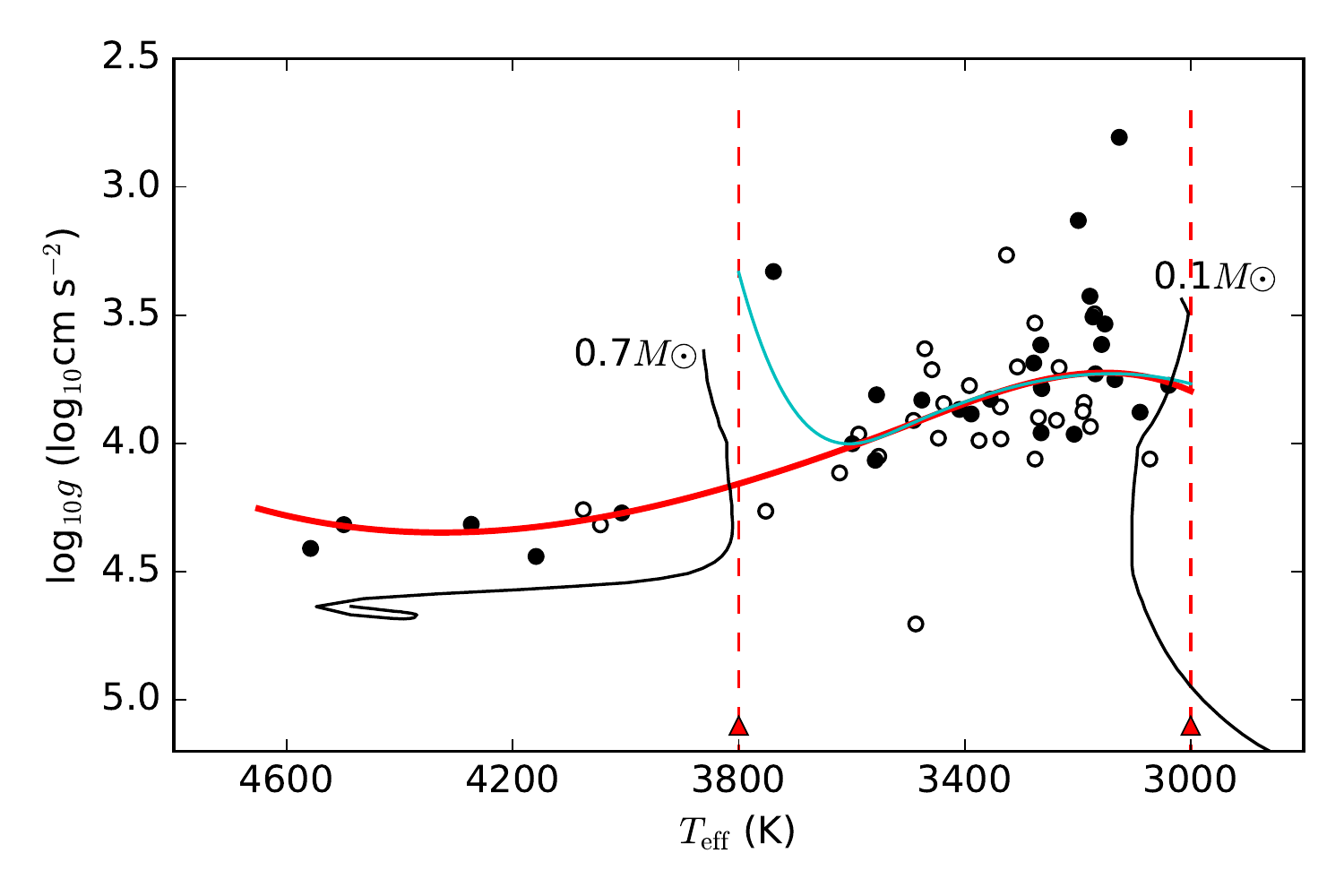}\\
\includegraphics[scale=0.58]{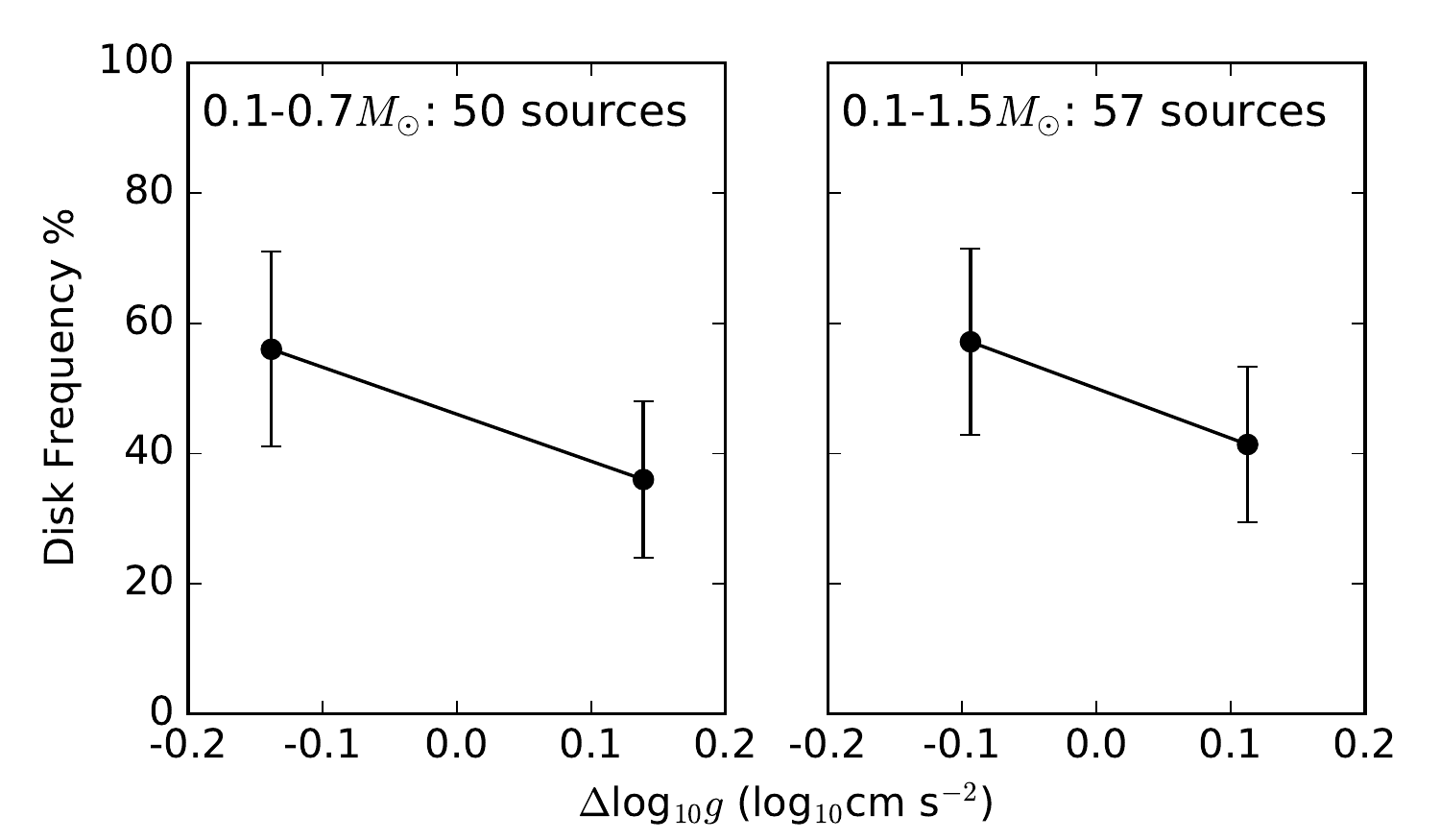}
\caption{Upper panel: Distribution of spectroscopic $T_{\rm eff}$ and log$g$ in NGC 1333 
(black dots for thick disks, and white dots for diskless stars). The
red and cyan solid lines are median surface gravity derived by fitting cubic spline 
functions to the median-filtered trend of log$g$ with $T_{\rm eff}$ (red: using 57 stars
with 4750 K$<T_{\rm eff}<3000$ K; cyan: using 50 stars with 3000K$<T_{\rm eff}<3800$K). 
The black lines are BCAH98 evolutionary tracks for 0.1 and 0.7 $M_{\odot}$ stars,
and their corresponding effective temperatures are marked by the dashed red line. 
Bottom panel: Disk frequency as a function of $\Delta \rm{log} \it g$ (Eq. 
\ref{eq:dellogg}). In each mass range,
the two bins of $\Delta \rm{log} \it g$ are divided to contain similar number of sources.
 \label{fig:HRD_NGC1333}}
\end{figure}

\subsection{NGC 1333}
The upper panel of Figure \ref{fig:HRD_NGC1333} shows the distribution of 57
sources in NGC 1333 on the log$g$--$T_{\rm eff}$ diagram. Disk-bearing  stars are shown
in black, and diskless stars are marked in white. We define 
\begin{equation}\label{eq:dellogg}
\Delta \rm{log} \it g_{i} = \rm{log} \it g_{i}- \rm{log} \it g_{\rm median,\it i},
\end{equation}
where log$g_{i}$ is the surface gravity of the $i$th star, and log$g_{\rm median,\it i}$ 
is the median surface gravity at the effective temperature of $T_{\rm eff, \it i}$. 
We empirically determine log$g_{\rm median}$ by fitting a cubic spline to the median-filtered 
trend of log$g$ with $T_{\rm eff}$ using stars on the log$g$--$T_{\rm eff}$
plane. The red solid line is the resulting log$g_{\rm median}$ fitted by using all of the 57 stars,
while the cyan line is log$g_{\rm median}$ fitted using the 50 YSOs between 3000 K and
3800 K.

At a given (relatively low) effective temperature, a younger PMS star 
has been through shorter time of gravitational contraction along the Hayashi line 
\citep{1961PASJ...13..450H}, and thus it typically has larger stellar radius and lower 
surface gravity, i.e., smaller $\Delta \rm{log} \it g$. However, if the effective 
temperature is relatively high (depending on stellar age and 
metallicity), a smaller $\Delta \rm{log} \it g$ may either indicate a younger age or a 
higher mass, because more massive stars ($M>0.7$ $M_{\odot}$) will not remain convective at later
stages of PMS evolution and would evolve towards higher $T_{\rm eff}$. 
The evolutionary tracks of a 0.1 $M_{\odot}$ star and a 0.7 $M_{\odot}$ star from the BCAH98 model
are also shown in the upper panel of Figure \ref{fig:HRD_NGC1333}. 
Their effective temperatures roughly correspond to
3000 K and 3800 K, which are found by looking at the intersections of the evolutionary tracks and
the red line. To investigate how disk frequency is dependent on stellar age, 
we first consider the 50 stars with 3000 K $<T_{\rm eff}<3800$ K, (0.1--0.7 $M_{\odot}$),
because $\Delta \rm{log} \it g$ is directly related to stellar age in this temperature (mass) range.
After that, we also look at the result from all of the 57 stars
(0.1--1.5 $M_{\odot}$) to see if the trend of disk frequency changes.
We adopt the red line as the median of surface gravity in all cases, since the cyan line
overfits the data at the boundary region ($\sim$3800 K).

\begin{figure}[ht!]
\includegraphics[width=\columnwidth]{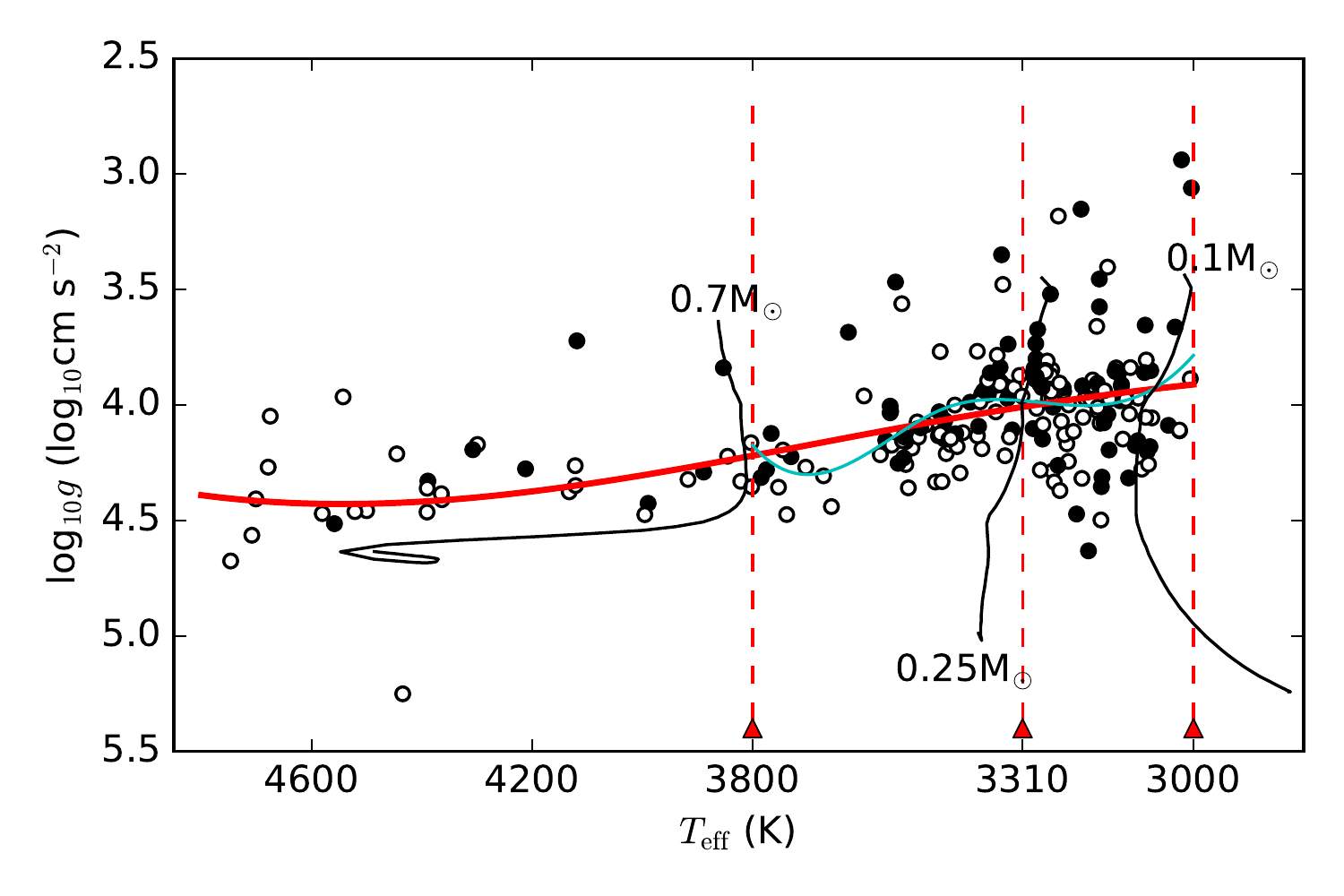}
\includegraphics[width=\columnwidth]{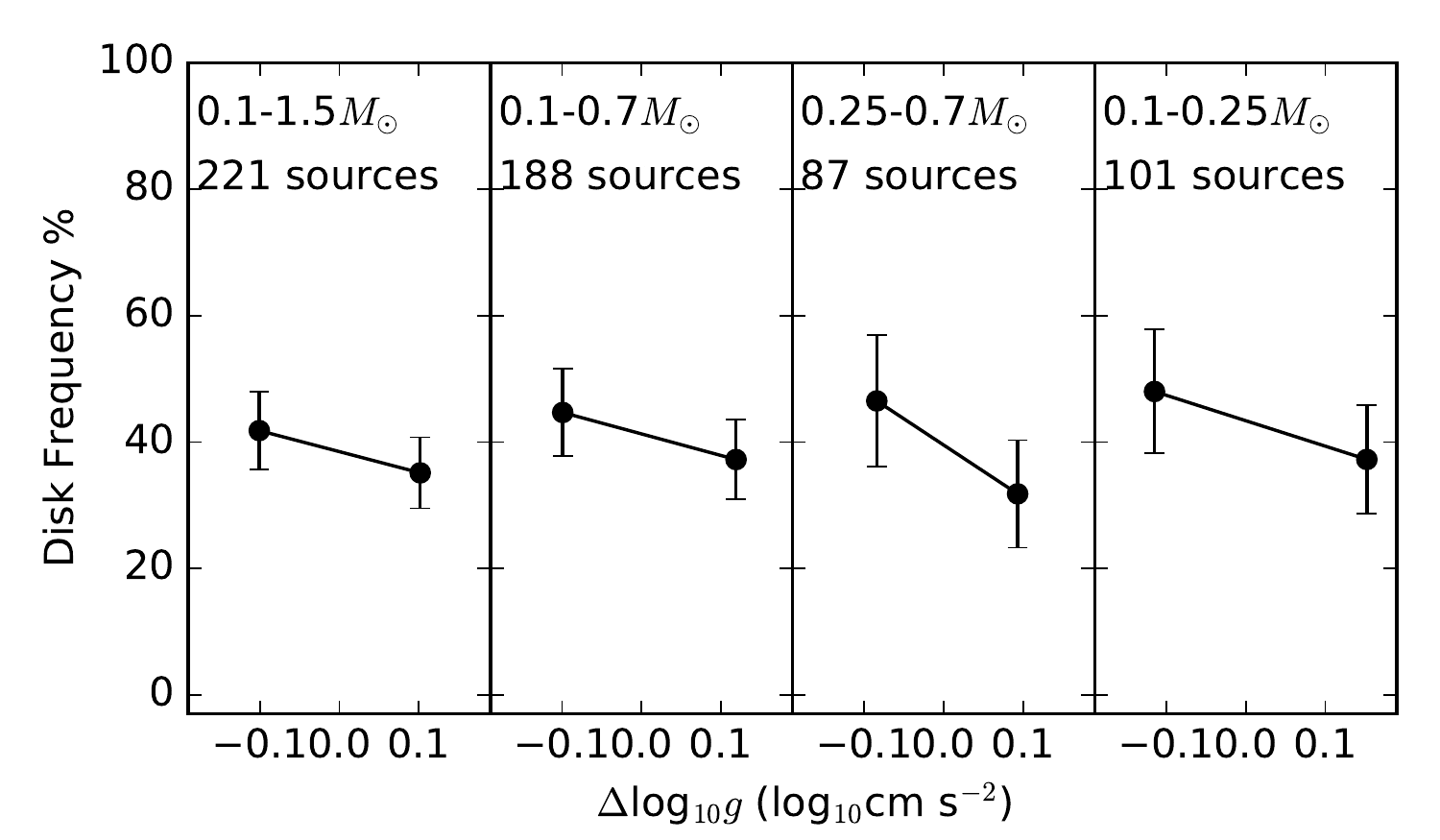}
\caption{Upper panel: Distribution of spectroscopic $T_{\rm eff}$ and log$g$ in IC 348 
Bottom panel: Disk frequency as a function of $\Delta 
\rm{log} \it g$ (Eq. \ref{eq:dellogg}). 
See Figure \ref{fig:HRD_NGC1333}.\label{fig:HRD_IC348}}
\end{figure}

\begin{figure*}[ht!]
\includegraphics[width=\textwidth]{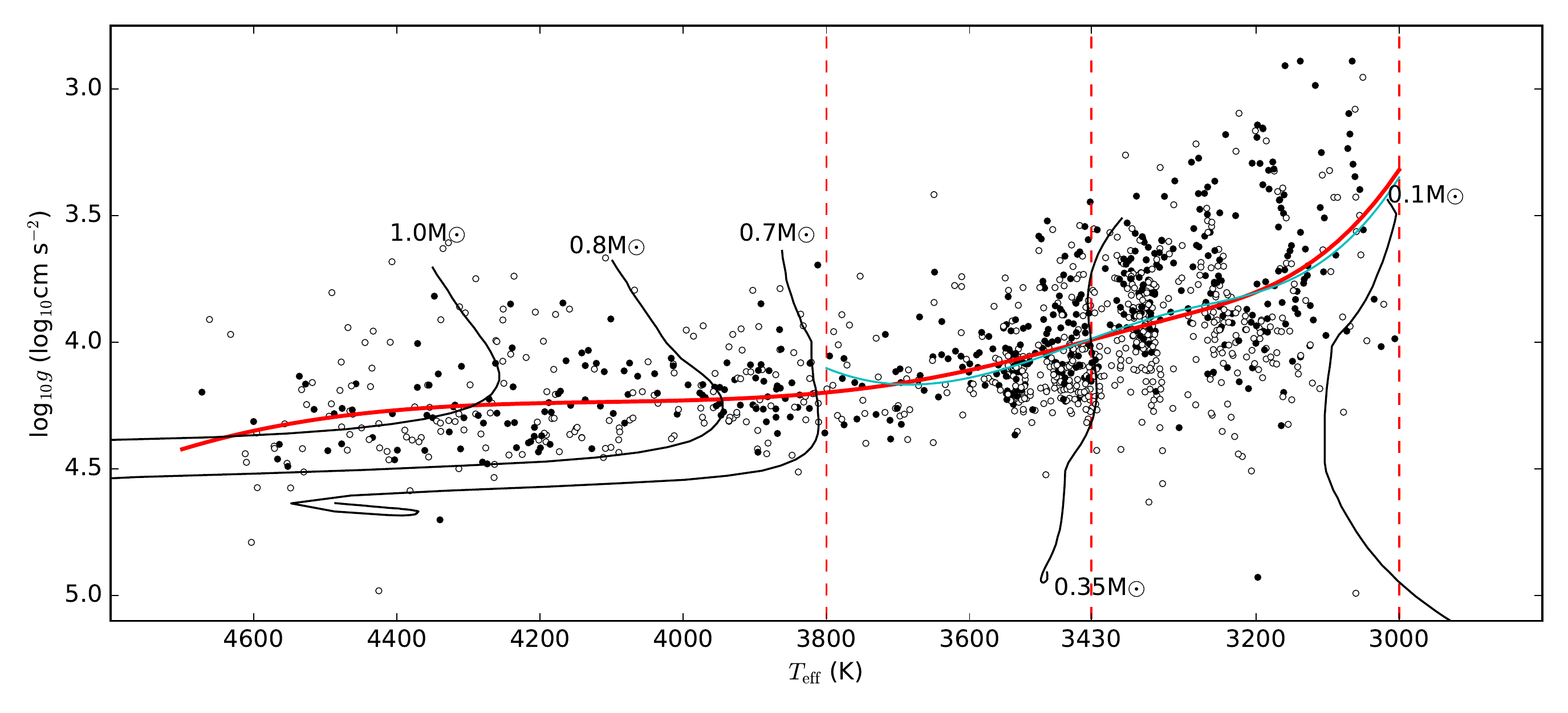}
\caption{Distribution of spectroscopic $T_{\rm eff}$ and log$g$ in Orion A. See Figure \ref{fig:HRD_NGC1333}.
\label{fig:HRD_OrionA}}
\end{figure*}

The lower panel of Figure \ref{fig:HRD_NGC1333} shows disk frequency as a function of
$\Delta \rm{log} \it g$. The two bins of $\Delta \rm{log} \it g$
are divided to contain the same number of sources.
In each case, there is a decrease of primordial disk frequency for older stars.
Including the 7 sources more massive than 0.7 $M_{\odot}$ does not change the general trend.
Unfortunately, we do not have enough sources in this region to derive a strong conclusion
that such a decrease in disk frequency indicates an evolution of primordial disks in NGC 1333. 

\subsection{IC 348}
The upper panel of Figure \ref{fig:HRD_IC348} shows the distribution of sources in IC 348 
on the log$g$--$T_{\rm eff}$ diagram. Over-plotted are the BCAH98 
evolutionary tracks of 0.1, 0.25, and 0.7 $M_{\odot}$ stars. The median surface gravity
is fitted in the same way as in the last section, and we adopt the red line as
log$g_{\rm median}$. Similar to what we have done in the last section,
we also individually look at disk frequency as a function of stellar age
in the 0.1--0.7 $M_{\odot}$ and 0.1--1.5 $M_{\odot}$ mass ranges. Moreover,
since our sample in IC 348 is representative down to 0.25 $M_{\odot}$, we also
make a cut at 0.25 $M_{\odot}$. This mass roughly corresponds to 3310 K,
which is found by looking at the intersection of the evolutionary track of a 0.25 $M_{\odot}$ star
and the red line. The result from 0.25--0.7 $M_{\odot}$ may be more convincing than that
from 0.1--0.25 $M_{\odot}$, since our sample is not representative in the latter mass range.

As is shown in the lower panel of Figure \ref{fig:HRD_IC348}, in all mass ranges we have 
seen a slight drop of disk frequency for relatively older stars. The drop is most significant
in the 0.25--0.7$M_{\odot}$ range. However, considering the statistical uncertainty, 
this is also consistent with the same disk frequency for different $\Delta \rm{log} \it g$ bins.

\subsection{Orion A}
Similar to the analysis in IC 348, the upper panel of Figure \ref{fig:HRD_OrionA} shows 
the distribution of sources in Orion A on the log$g$--$T_{\rm eff}$ diagram. 
All of the 1200 sources have $A_J<3.0$. Since the Orion A sample is representative
down to $\sim$0.35 $M_{\odot}$ at this extinction limit, we make a cut at 0.35 $M_{\odot}$,
which roughly corresponds to 3430 K. The red and cyan lines have the same meaning
as in the last section, and we adopt the red as log$g_{\rm median}$.

In the lower panel of Figure \ref{fig:HRD_OrionA}, we show disk frequency as a function of 
$\Delta {\rm log}g$ in the mass ranges of 0.1--1.5 $M_{\odot}$, 0.7--1.5 $M_{\odot}$,
0.35--0.7 $M_{\odot}$, and 0.1--0.35 $M_{\odot}$. In each mass range, we divide the sample
according to $\Delta {\rm log}g$ into three bins with equal number of stars.
The definite decrease of disk frequency from (55.6$\pm$6.7)\% to (41.1$\pm$5.8)\%,
and finally down to (27.4$\pm$4.7)\% in 0.35--0.7 $M_{\odot}$ strongly
indicates that disk evolution happens, because our sample is representative in this mass range.
The absolute disk frequency for 0.1--0.35 $M_{\odot}$ suffers from observational bias. 
However, since log$g$ and IR excess are two independent measurements, 
we may not expect that such a bias depends on stellar age. 
Therefore, the relative disk frequency in the mass range of 0.1--0.35 $M_{\odot}$
with different $\Delta$log$g$ (from (55.5$\pm$5.7)\% to (41.9$\pm$4.9)\%, to (23.8$\pm$3.7)\%) may still 
provide evidence of disk evolution for the lowest mass stars. 

The trend of disk frequency in the mass bin of 0.7--1.5 $M_{\odot}$ 
is more complicated than other bins --- it first goes up, and then goes down as $\Delta {\rm log}g$ increases. 
As mentioned before, in this mass range, an increase in
$\Delta {\rm log}g$ may indicate older age or lower mass (or both). 
If the higher $\Delta {\rm log}g$ originates from lower mass, 
then disk frequency should increase for higher $\Delta {\rm log}g$, because 
Section \ref{subsec:mass_OrionA} and Figure 
\ref{fig:freq_OrionA} show that disk frequency increases from 1.5 to 0.7 $M_{\odot}$.
On the other hand, if stellar age is the dominant factor of the variation of surface gravity,
then disk frequency should decrease towards higher $\Delta {\rm log}g$.
We may need to combine both factors to explain our result.

\begin{figure}[ht!]
\includegraphics[width=\columnwidth]{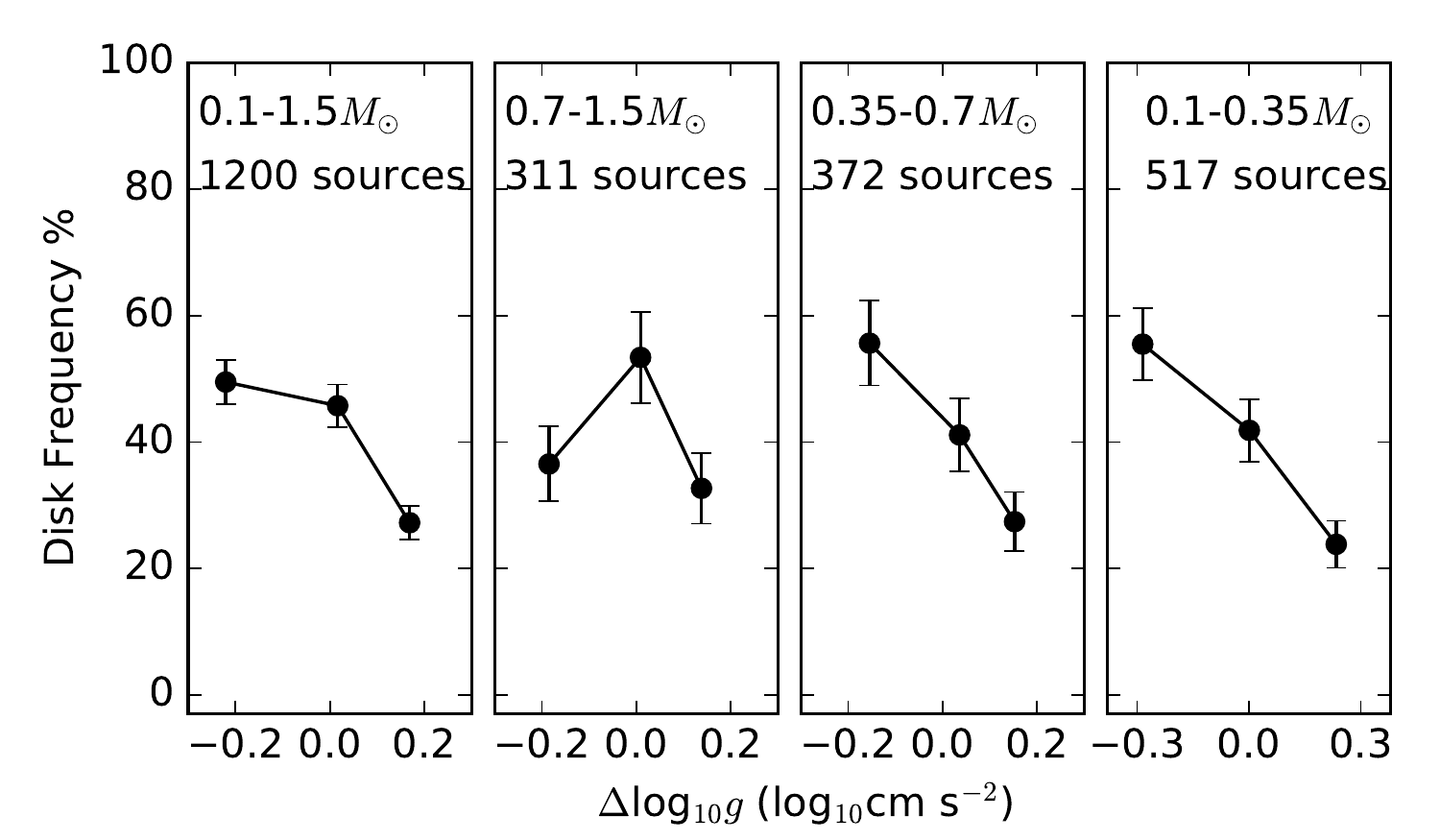}
\caption{Disk frequency as a function of $\Delta \rm{log} \it g$ (Eq. \ref{eq:dellogg}). 
\label{fig:age_OrionA}}
\end{figure}

Combining all stars from 0.1 to 1.5 $M_{\odot}$ and dividing them into three bins
of $\Delta {\rm log}g$, we find that disk frequency decreases from 
(49.5$\pm$3.5)\% to (45.8$\pm$3.4)\%, and then to (27.3$\pm$2.6)\%. The large number of sources in this cluster 
as well as its intrinsic spread of age allow us to draw robust conclusion of 
the detection of intra-cluster disk evolution.


\section{Summary}\label{sec:conclusion}
Utilizing a sample of stars observed by the IN-SYNC project
with accurately-derived stellar parameters, we investigate disk evolution in three young clusters.
First, in each cluster, averaged over the spread of age, we carefully studied
how disk lifetime is dependent on stellar mass for low-mass stars.
Previous results show that disk lifetime for intermediate-mass stars is much shorter than 
that for low-mass stars. This is consistent with our result.
However, we think disk lifetime for most stars in the low-mass range 
are similar in the sense that we do not observe a significant increase of disk 
frequency from 0.8 to 0.1 $M_{\odot}$.

Second, in each cluster, within a certain mass range,
disk frequency (as calculated by the fraction of stars with 4.5 $\micron$ excess)
for stars with larger log$g$ is lower than that with smaller
log$g$, indicating a longer time of disk dispersal for older stars.
The result from Orion A is strong and the most prominent.
Evidence of intra-cluster primordial disk evolution in NGC 1333 and IC 348 is still tentative,
since the derived frequencies are all compatible within 1-$\sigma$.
Spectroscopic measurements of $T_{\rm eff}$ and log$g$ for more low-mass
stars in the two clusters are needed in the future to draw robust evidence
of intra-cluster disk evolution. We have demonstrated in this paper the usage of log$g$
as an age indicator for low mass stars in the study of disk evolution. This methodology 
can be applied to other young clusters in the future.

\appendix \label{appendix}
In Table \ref{tab:tablesplit}, we provide information of the 1641 sources used in our sample, 
including 70 in NGC 1333, 239 in IC 34, and 1332 in the Orion A molecular cloud. 
Apart from stellar parameters and photometric data, we also indicate whether 
hydrogen emission lines in the brackett series ($n'=4$) can be observed in their $H$ band APOGEE spectra.
Whenever brackett emission lines can be identified, we also present the equivalent width (EW), line flux, and full width half maximum (FWHM) of
the Br11 line at 1681nm -- the strongest Brackett line in the APOGEE wavelength range.

It has been known for long that hydrogen emission lines are indicative of large-scale gas flows.
Generally speaking, the profiles of Br11 show complex behaviors: most exhibit lines that are symmetric about line center;
while some show significant blueward asymmetry or redshifted absorption (e.g. 2M03292187$+$3115363 in NGC 1333), 
which are direct evidence of mass infall \citep{1994AJ....108.1056E};
a few stars (e.g. 2M05341221$-$0450072 in Orion A) have overlying absorption on top of the emission profile.
Just like other high-excitation hydrogen emission lines \citep{1997IAUS..182P.272F, 1998ApJ...492..743M}, 
blueshifted absorption commonly seen in H$\alpha$ can seldom be observed.

In the densest region of ONC, stellar spectra are contaminated by narrow (FWHM $<70$ km s$^{-1}$) nebula emission due to imperfect sky subtraction;
However, in our sample, the Br11 emission line with large line widths (FWHM $>100$ km s$^{-1}$) are only found in stars exhibiting infrared 
excess. 
Models with infalling gas via magnetospheric accretion have successfully reproduced these 
characteristics \citep{1998AJ....116..455M, 2001ApJ...550..944M}.
Others may in the future would like to model these line profiles.

\begin{table*}[htbp!]
\renewcommand{\thetable}{\arabic{table}}
\caption{1875 sources used in the sample\label{tab:tablesplit}}
\begin{splittabular}{ccccccccBcccccccc}
\hline
\hline
2MASS designation 	& ra 	\tablenotemark{a}		& dec  \tablenotemark{a}	& $T_{\rm eff}$		&log$g$			&$J$	 			 &$H$			  &$K_{\rm s}$		  
&$A_{J}$	&[4.5] 				&ditype\tablenotemark{b}	&brackett		&EW(Br11)		&Line Flux		&FWHM(Br11)	&Cluster\\
	    			& ($\degree$) 	&  ($\degree$) 	&( K)				&				&(mag)			 &(mag)			  &(mag)			  &(mag)				
				&(mag)		&			&		&	($\AA$)					& (erg s$^{-1}$ cm$^{-2}$)	&(km s$^{-1}$)&\\
\hline	  
03281101$+$3117292  & 52.04590	&31.29146	& 3374.6$\pm$8.6	& 3.99$\pm$0.03	& 12.441$\pm$0.021 & 11.457$\pm$0.028 & 11.028$\pm$0.024 
&0.874	& 10.692$\pm$0.055 	&0 		&0			&				&					&	&NGC 1333\\
03283651$+$3119289  & 52.15215	&31.32471 	& 3269.5$\pm$6.2	& 3.90$\pm$0.04	& 12.849$\pm$0.020 & 12.128$\pm$0.026 & 11.856$\pm$0.023 
&0.214	& 11.566$\pm$0.054	&0		&0			&				&					&	&NGC 1333\\
03292187$+$3115363  & 52.34114       &31.26008	& 4158.7$\pm$8.7	& 4.44$\pm$0.02      & 11.176$\pm$0.019 & 10.151$\pm$0.028 &    9.504$\pm$0.018 
&1.053	&   8.197$\pm$0.057 	&1		&1			&-1.22$\pm0.02$	&1252.01$\pm$21.65	&283.65$\pm$6.47	&NGC 1333\\
05344929$-$0518555 & 83.70540 	& $-$5.31542 	& 3459.5$\pm$10.7	& 4.09$\pm$0.04 & 12.082$\pm$ 0.019 & 11.367 $\pm$0.03 & 11.158$\pm$0.019 
& 0.128	& 10.952$\pm$0.052 	& 0 		& 1 			& -0.18$\pm$0.05 	& 59.38$\pm$15.92 		& 27.56$\pm$1.83 	& Orion A\\
05332136$-$0521347 	& 83.33903 & $-$5.35964 	& 4276.0$\pm$200.0 	& 4.28$\pm$0.35 	& 11.253$\pm$0.019 & 10.332$\pm$0.021 & 9.769$\pm$0.017 
& 0.838 	& 8.726$\pm$0.051		 &1 		& 1 			& -1.05$\pm$0.04 	& 867.12$\pm$28.14		& 265.29$\pm$38.15 & Orion A\\
\hline
\end{splittabular}
\tablenotetext{a}{Right ascension and declination given in the 2MASS Point Source Catalog; epoch in J2000.}
\tablenotetext{b}{Disk type. 0: diskless star; 1: disk-bearing star.}
\tablecomments{This table is available in its entirety in machine-readable form.}
\end{table*}

\acknowledgments
We acknowledge the anonymous referee, Lee Hartmann, Nuria Calvet, and Lynne Hillenbrand for valuable suggestions that greatly improved this manuscript.
Stellar parameters for the Perseus sources are derived by Michiel Cottaar. 
Jonathan B. Foster also contributed to the design of the IN-SYNC program. 
Special thanks to Jesus Hern{\'a}ndez and Mark J. Pecaut for patiently answering questions about details 
and techniques in their work. Yuhan Yao acknowledge China Scholarship Council (CSC) for supporting this research.

\end{document}